\documentclass[aps,onecolumn,prx]{revtex4-1}
\usepackage{graphicx,amsmath, amsfonts,bm, color}

\usepackage{xcolor,hyperref}
\hypersetup{
   colorlinks,
   linkcolor={blue!50!black},
   citecolor={blue!50!black},
   urlcolor={blue!80!black}
}

\usepackage{enumitem}
\usepackage{bm}
\usepackage{amssymb}
\usepackage{wasysym}
\usepackage{caption}
\newcommand{\sgn}{\text{sgn}}
\newcommand{\pa}{\partial}

\renewcommand{\=}{\!=\!}

\newcommand{\tel}{\tau^{\rm el}}

\begin{document}

\title{Spatiotemporal dynamics of frictional systems:\\ The interplay of interfacial friction and bulk elasticity}
\author{Yohai Bar-Sinai$^{1,2}$}
\author{Michael Aldam$^{1}$}
\author{Robert Spatschek $^{3}$}
\author{Efim A.~Brener$^{3,4}$}
\author{Eran Bouchbinder$^{1}$}
\thanks{eran.bouchbinder@weizmann.ac.il}
\affiliation{
$^{1}$Chemical and Biological Physics Department, Weizmann Institute of Science, Rehovot 7610001, Israel\\
$^{2}$John A.~Paulson School of Engineering and Applied Sciences, Harvard University, Cambridge, MA 02138, USA\\
$^{3}$Institute for Energy and Climate Research, Forschungszentrum J\"ulich, D-52425 J\"ulich, Germany\\
$^{4}$Peter Gr\"unberg Institut, Forschungszentrum J\"ulich, D-52425 J\"ulich, Germany\\}

\begin{abstract}
Frictional interfaces are abundant in natural and engineering systems, and predicting their behavior still poses challenges of prime scientific and technological importance. At the heart of these challenges lies the inherent coupling between the interfacial constitutive relation --- the macroscopic friction law --- and the bulk elasticity of the bodies that form the frictional interface. In this feature paper, we discuss the generic properties of the macroscopic friction law and the many ways in which its coupling to bulk elasticity gives rise to rich spatiotemporal frictional dynamics. We first present the widely used rate-and-state friction constitutive framework, discuss its power and limitations, and propose extensions that are supported by experimental data. We then discuss how bulk elasticity couples different parts of the interface, and how the range and nature of this interaction are affected by the system's geometry. Finally, in light of the coupling between interfacial and bulk physics, we discuss basic phenomena in spatially-extended frictional systems, including the stability of homogeneous sliding, the onset of sliding motion and a wide variety of propagating frictional modes (e.g.~rupture fronts, healing fronts and slip pulses). Overall, the results presented and discussed in this feature paper highlight the inseparable roles played by interfacial and bulk physics in spatially-extended frictional systems.
\end{abstract}

\maketitle

\section{Introduction}

\begin{figure}[ht]
  \centering
  \includegraphics[width=.7\textwidth]{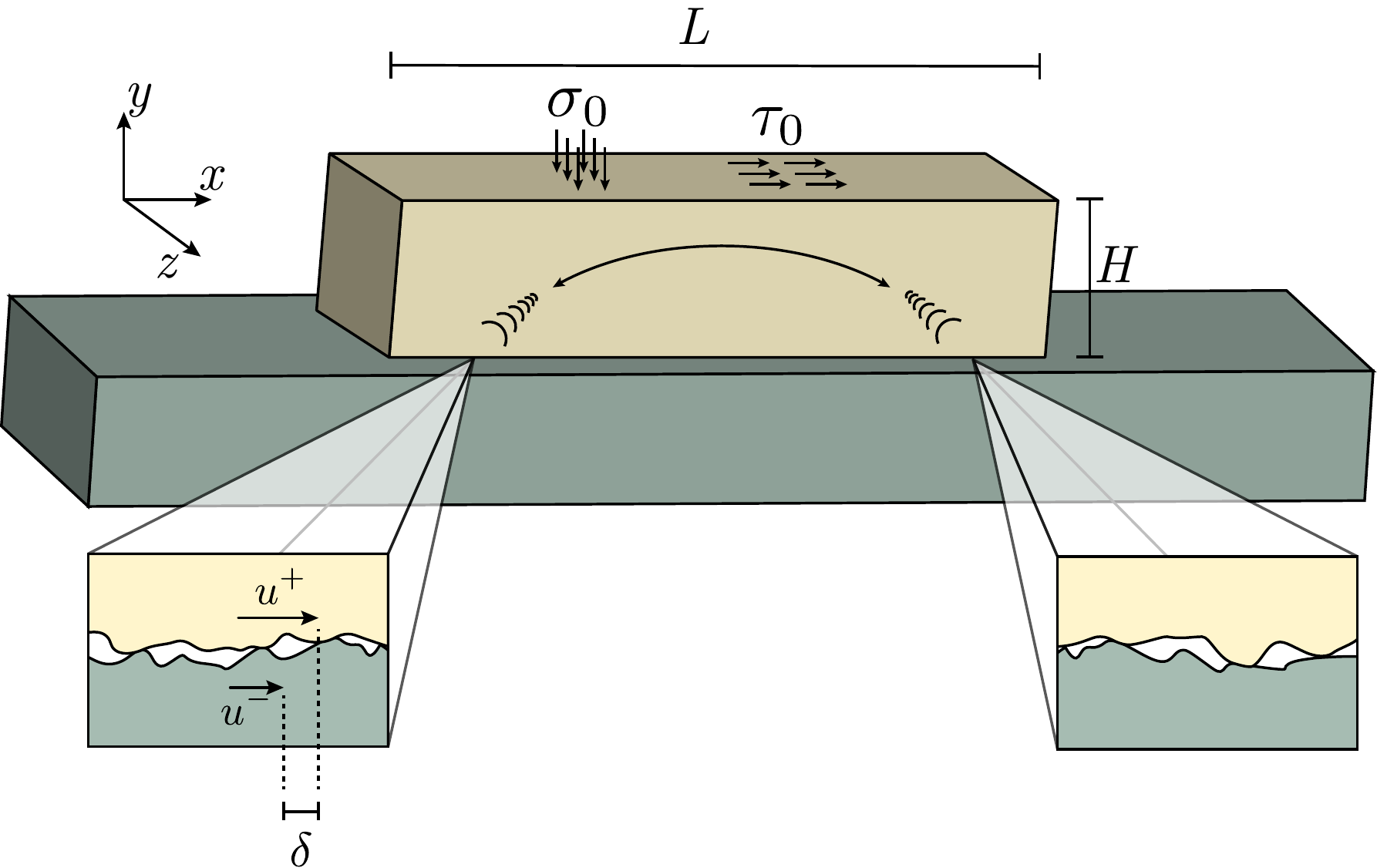}
  \caption{A schematic of a frictional system. Two macroscopic bodies are in contact, the interface between them is composed of a discrete collection of contact asperities (see zoom in). Here we plot the general case in which the bodies are made of different materials (hence the different colors used) and have different geometries. The system is loaded normally and tangentially ($\sigma_0$ and $\tau_0$, respectively) far from the interface. Stress conditions at the interface, which give rise to frictional motion, result from the interplay between local frictional rheology (see zoom in) and forces mediated by the elastic bulks, induced by the loading and by differential sliding in distant parts of the interface (see curved double-arrow). The height $H$ and length $L$ of one of the bodies are marked and a coordinate system is introduced. The local slip displacement $\delta$, the difference between the local displacement of the upper/lower bodies ($u^{\pm}$), is also schematically denoted.}
  \label{fig:sketch}
\end{figure}

The dynamics of frictional systems have been the focus of intense scientific research in the last few decades~\cite{Bowden1950,Persson1998,Svetlizky2019}, with applications ranging from atomic and nano-scale systems~\cite{Sorensen1996, Li2011, Vanossi2013}, through engineering and control systems~\cite{Armstrong-Helouvry1994,Wojewoda2008}, to geophysical systems~\cite{Marone1998a,Ben-Zion2001, Scholz2002, Ohnaka2013}. Despite the progress made in understanding and controlling friction at different scales~\cite{Gnecco2001, Baumberger2006, Ben-Zion2008, Vakis2018}, various basic aspects of frictional systems are not yet well-understood. As frictional systems and processes span a wide range of spatiotemporal scales, the problem can be approached from different perspectives. The focus of the present feature paper is on spatially-extended frictional systems approached from a macroscopic (continuum) perspective. Our major goal is to discuss the generic properties of macroscopic interfacial constitutive relations, within the rate-and-state friction framework~\cite{Dieterich1978,Dieterich1979,Rice1983,Ruina1983,Heslot1994,Marone1998a,Baumberger2006}, and to demonstrate that the overall dynamics of frictional systems emerge from the inherent coupling between interfacial tribology and the bulk elasticity of the bodies forming the interface. As such, the physics and dynamics of frictional systems are highlighted to result from the inherent interplay between the local, micro-scale tribological behaviour of frictional interfaces and the non-local, long-range interaction between different parts of the interfaces, which are mediated by macro-scale bulk elasticity (cf.~Fig.~\ref{fig:sketch}).

To see the nature and origin of this inherent coupling let us consider two bodies in frictional contact and the associated slip displacement $\delta$, which is the difference between the displacement of the upper and lower bodies $u^\pm$, respectively, at the interface (cf.~Fig.~\ref{fig:sketch}). In spatially-extended frictional systems, $\delta$ generically varies along the interface, and hence should be in fact represented by a field $\delta(x,t)$, where $x$ is the coordinate along the interface that is located at $y\=0$ and $t$ is the time (cf.~Fig.~\ref{fig:sketch} and note that we do not consider the $z$ direction, which is nevertheless marked in the figure). A coarse-grained constitutive law relates the frictional stress $\tau(x,t)$ (sometimes termed the frictional strength/resistance) to quantities such as the slip displacement field $\delta(x,t)$, the slip velocity field $v(x,t)\=\partial_t\delta(x,t)$, the interfacial normal (compressive) stress $\sigma(x,t)$ and possibly other interfacial fields, and can be expressed as
\begin{equation}
\tau(x,t) = f\!\left[\delta(x,t), \partial_t\delta(x,t), ...\right]\,\sigma(x,t) \ .
\label{eq:strength}
\end{equation}
Here the dots represent yet unspecified internal state fields, which satisfy their own evolution equations. The latter, together with the functional $f[\cdot]$ that appears as a proportionality factor (``friction coefficient'') between $\sigma(x,t)$ and $\tau(x,t)$, define the coarse-grained constitutive relation. The purely local nature of $f[\cdot]$ is manifested by the absence of spatial derivatives of $\delta(x,t)$ in it. When bulk elasticity is not taken into account, i.e.~when the bulks are considered infinitely rigid as in spring-block models~\cite{Heslot1994,Baumberger2006,Popov2010}, Eq.~\eqref{eq:strength} takes the form $\tau_0\=f\,\sigma_0$, where $\tau_0$ and $\sigma_0$ are the far-field applied shear stress and normal (compressive) stress, respectively (cf.~Fig.~\ref{fig:sketch}). In this case, the frictional problem has a single degree of freedom, namely $\delta(t)$. A major goal of this feature paper is to understand how spatial degrees of freedom, emerging from bulk elasticity, change/extend the relation $\tau_0\=f\,\sigma_0$ and to demonstrate their far-reaching implications for frictional dynamics.

When the elasticity of the bodies forming the interface is taken into account, Eq.~\eqref{eq:strength} serves as a boundary condition that relates two components of the stress tensor field $\sigma_{ij}(x,y,t)$ as the interface is approached, $y\!\to\!0$. In particular, $\tau(x,t)$ equals the bulk shear stress, $\sigma_{xy}(x,y,t)$, and $\sigma(x,t)$ equals the bulk normal (compressive) stress, $-\sigma_{yy}(x,y,t)$, in the limit $y\!\to\!0$. The interfacial shear stress and normal (compressive) stress, $\sigma_{xy}(x,y\=0,t)$ and $-\sigma_{yy}(x,y\=0,t)$, depend on the far-field applied shear stress and normal (compressive) stress, $\tau_0$ and $\sigma_0$ respectively, {\bf and} on stresses that emerge from all processes that are mediated by the presence of the bulk of the bodies forming the interface. The latter depend on gradients of the slip displacement field, $\partial_x\delta(x,t)$, and on the slip velocity field, $v(x,t)\=\partial_t\delta(x,t)$, and can be expressed as ${\cal F}_{_{\!\tau}}\!\left[\partial_x\delta(x,t), \partial_t\delta(x,t)\right]$ and ${\cal F}_{_{\!\sigma}}\!\left[\partial_x\delta(x,t), \partial_t\delta(x,t)\right]$, respectively. In case the relevant bulk physics is faithfully represented by linear elastodynamics, as is widely assumed elsewhere and as is also adopted here, the interfacial stresses take the form
\begin{equation}
\sigma_{xy}(x,y\=0,t) = \tau_0 - {\cal F}_{_{\!\tau}}\!\left[\partial_x\delta(x,t), \partial_t\delta(x,t)\right],~~~~ -\sigma_{yy}(x,y\=0,t) = \sigma_0 - {\cal F}_{_{\!\sigma}}\!\left[\partial_x\delta(x,t), \partial_t\delta(x,t)\right] \ .
\label{eq:elastodynamics}
\end{equation}
The linear functionals ${\cal F}_{_{\!\tau}}\!\left[\partial_x\delta(x,t), \partial_t\delta(x,t)\right]$ and ${\cal F}_{_{\!\sigma}}\!\left[\partial_x\delta(x,t), \partial_t\delta(x,t)\right]$ generically describe long-range spatiotemporal interactions (i.e.~note the explicit appearance of spatial gradient in $\partial_x \delta(x,t)$, which schematically represents also higher order spatial derivatives and even a spatial integral) and interface-bulk radiation effects. They are affected by dimensionality, by the geometry of the bodies forming the interface and by their constitutive parameters (cf.~Fig.~\ref{fig:sketch}).

Using the equalities $\tau(x,t)\=\sigma_{xy}(x,y\=0,t)$ and $\sigma(x,t)\=-\sigma_{yy}(x,y\=0,t)$, Eq.~\eqref{eq:strength} takes the form
\begin{equation}
\tau_0-{\cal F}_{_{\!\tau}}\!\left[\partial_x\delta(x,t), \partial_t\delta(x,t)\right] = f\!\left[\delta(x,t), \partial_t\delta(x,t), ...\right] \left(\sigma_0 - {\cal F}_{_{\!\sigma}}\!\left[\partial_x\delta(x,t), \partial_t\delta(x,t)\right] \right) \ ,
\label{eq:coupling}
\end{equation}
which sets the framework for the discussion in this paper. This basic equation clearly demonstrates that frictional dynamics, which correspond to solutions for $\delta(x,t)$, inherently depend on both the interfacial constitutive relation $f\!\left[\delta(x,t), \partial_t\delta(x,t), ...\right]$ (along with the internal state fields and their own evolution equations, to be discussed below), and on the bulk constitutive contributions ${\cal F}_{_{\!\tau}}\!\left[\partial_x\delta(x,t), \partial_t\delta(x,t)\right]$ and ${\cal F}_{_{\!\sigma}}\!\left[\partial_x\delta(x,t), \partial_t\delta(x,t)\right]$. Consequently, the dynamics of frictional systems inseparably emerge from the coupling between interfacial and bulk physics, as encapsulated in Eq.~\eqref{eq:coupling}.

In the following parts of the paper we first discuss the generic properties of the interfacial constitutive relation $f\!\left[\delta(x,t), \partial_t\delta(x,t), ...\right]$,  including internal state fields and their evolution equations, and then consider the bulk constitutive contributions ${\cal F}_{_{\!\tau}}\!\left[\partial_x\delta(x,t), \partial_t\delta(x,t)\right]$ and ${\cal F}_{_{\!\sigma}}\!\left[\partial_x\delta(x,t), \partial_t\delta(x,t)\right]$, for various dimensionalities, physical situations and geometric configurations relevant for a broad range of frictional systems. While in so doing we build on extensive effort and progress accumulated in the literature over the last few decades, we are also strongly biased toward our own recent work. As such, this paper inevitably presents our personal perspective on the issues at stake. Yet, we hope that the emerging physical picture is sufficiently broad and general to be useful for scientists interested in a wide variety of frictional systems. Whenever possible, we make an effort to highlight basic concepts and generic physical effects, while minimizing the technical complexity of the presentation and referring the readers to a more technically involved literature.

\section{Generic properties of interfacial constitutive laws}
\label{sec:constitutive}

A widely-used class of interfacial constitutive laws, termed \textit{rate-and-state friction} (RSF) models, originated from the seminal works of Dieterich~\cite{Dieterich1978,Dieterich1979,Dieterich1986} and Rice and Ruina \cite{Rice1980a,Ruina1983} in the late 1970's, which elaborated on ideas formulated by Bowden and Tabor~\cite{Bowden1950} and Rabinowicz \cite{Rabinowicz1951} in the 1950's. These models introduce one or more coarse-grained internal state variables to account for the evolution of the interfacial structural state (and hence they carry memory of the interfacial history), and rate dependence to account for interfacial rheology. In this section, we will briefly describe the conventional RSF model, discuss what we view as its power and possible shortcomings, and consequently propose modifications. Our ultimate goal is to provide an overview of the generic properties of interfacial constitutive laws.

\subsection{Conventional RSF models}
\label{sec:RSF}

It is now well-established that a crucial physical quantity affecting the dynamics of spatially-extended dry frictional interfaces is the real contact area $A_r$, which is typically orders of magnitude smaller than the nominal (apparent) contact area $A_n$ \cite{Dieterich1994a, Rubinstein2004, Baumberger2006, Rubinstein2007, Nagata2008}. That is, a spatially-extended interface separating two macroscopic bodies in frictional contact is composed of a collection of discrete contact asperities (i.e.~it is a multi-contact interface, cf.~Fig.~\ref{fig:sketch}) whose total area is typically much smaller than the nominal overlap of the bodies. A major implication of this physical picture is that the frictional stress $\tau$ can be written, following Bowden and Tabor~\cite{Bowden1950}, as $\tau\=(A_r/A_n) \tau_s(v)$, where $\tau$ is the frictional stress of Eq.~\eqref{eq:strength}, $\tau_s(v)$ is a rate-dependent asperity-level shear stress and $v$ is the interfacial slip velocity. The normalized real contact area, $A\!\equiv\!A_r/A_n$, is commonly expressed in terms of another state variable $\phi$, of time dimensions, as
\begin{equation}
 A(\phi)\equiv\!A_r/A_n=\frac{\sigma}{\sigma_{\hbox{\tiny H}}}\left[1+b\log\left(\frac{\phi}{\phi^*}\right)\right]\ .
\label{eq:Awrong}
\end{equation}
Here $\sigma$ is the interfacial normal stress of Eq.~\eqref{eq:strength}, $\sigma_{\hbox{\tiny H}}$ is the material hardness \cite{Estrin1996,Baumberger2006,Rubinstein2004} and $b$ is a dimensionless material parameter, typically of the order of $10^{-2}$. In this formulation, $\phi^*$ is a rather arbitrary time scale required for dimensional consistency. Below we discuss this point in detail.

Equation \eqref{eq:Awrong} properly describes the outcome of a broad range of experimental works that have demonstrated that the real contact area (or the static frictional resistance, which is proportional to $A$ according to the Bowden and Tabor relation $\tau\=A(\phi)\tau_s(v)$) grows logarithmically with the time of contact in the absence of sliding~\cite{Dieterich1972, Beeler1994, Berthoud1999, Bureau2001, Rubinstein2004, Ben-David2010}. This is the well-known logarithmic aging of the static frictional resistance.
Thus, Eq.~\eqref{eq:Awrong} allows one to write the simple relation $\pa_t\phi\=1$ in the absence of sliding, $v\=0$.
Within this physical interpretation, $\phi$ is the typical lifetime of a contact, and therefore, if the typical slip distance required to destroy an ensemble of contacts is $D$, we expect $\phi(v)\=D/|v|$ under persistent (steady-state) sliding, $v\!\ne\!0$. This implies that $A$ decreases logarithmically with increasing sliding velocity $v$, in agreement with the prevalent observation that the steady-state frictional resistance (to be discussed below), for a broad range of velocities, decreases logarithmically with $v$~\cite{Dieterich1978, Tullis1986, Kilgore1993, Baumberger1999, Reches2010}. To actually show this, one should consider $\tau_s(v)$, which we do next.

The rate-dependent contribution to the frictional resistance, $\tau_s(v)$, is usually associated with a stress-biased thermal activation process~\cite{Baumberger1999, Rice2001, Nakatani2001, Baumberger2006, Beeler2007, Bar-Sinai2014}. Within a simplified model of a single energy barrier $E$, linearly biased by the applied stress according to $E\!=\!E_0-\Omega\,\tau$, where $\Omega$ is an activation volume and $E_0$ is the bare energy barrier, $\tau_s(v)$ takes the form~\cite{Rice2001, Bar-Sinai2014}
\begin{equation}
\tau_s(v)=\frac{k_B T}{\Omega}\sinh^{-1}\left(\frac{v}{2\,v_c} \exp\left[\frac{E_0}{k_B T}\right]\right)
\label{eq:tausinh} \ .
\end{equation}
Here $k_B$ is Bolzmann's constant, $T$ is the temperature and $v_c$ is a velocity scale setting the upper limit for the validity of the thermal activation model.
Combining Eqs.~\eqref{eq:Awrong} and \eqref{eq:tausinh}, we obtain
\begin{equation}
\frac{\tau(\phi,v)}{\sigma}=\frac{k_B T}{\Omega\,\sigma_{\hbox{\tiny H}}}\left[1\!+\!b\log\left(\frac{\phi}{\phi^*}\right)\right]\sinh^{-1}\!\!\left(\frac{v}{2\,v_c}\,\exp\left[\frac{E_0}{k_B T}\right]\right)
\label{eq:RSFsinh} \ .
\end{equation}
In the limit where $E_0\gg k_BT$, which is quite typical, it can be shown that this equation takes the well-known form of RSF friction~\cite{Rice2001, Bar-Sinai2014}
\begin{equation}
f(\phi,v) \simeq f_0+\alpha \log\left(\frac{v}{v_c}\right)+\beta\log\left(\frac{\phi}{\phi^*}\right)\ ,
\label{eq:RSF}
\end{equation}
where $f(\phi,v)\=\tau(\phi,v)/\sigma$ as in Eq.~\eqref{eq:strength}, higher order logarithmic terms have been omitted, and the dimensionless parameters $f_0$, $\alpha$ and $\beta$ can be expressed in terms of physical parameters defined above (see~\hyperref[sec:math_friction]{Appendix A}).

The second logarithmic contribution on the right hand side of Eq.~\eqref{eq:RSF} emerges from the evolution of the real contact area, i.e.~the state-dependent contribution of Eq.~\eqref{eq:Awrong}. The first logarithmic contribution, on the other hand, emerges from the interfacial rheology, i.e.~the rate-dependent contribution of Eq.~\eqref{eq:tausinh}. This logarithmic $v$-dependent behavior is directly supported by experiments in which the slip velocity $v$ changes abruptly, such that $\phi$ does not change appreciably, and the change in the frictional stress is measured on a short time scale. Under steady-state conditions, $\phi$ is slaved to the slip velocity according to $\phi\=D/|v|$, and the overall rate-dependence of $f(\phi\=D/|v|,v)\=f_{\rm ss}(v)$ is proportional to $\log(v)$, as mentioned above. This logarithmic dependence is velocity-weakening for $\alpha\!<\!\beta$ and velocity-strengthening for $\alpha\!>\!\beta$. This issue will be further discussed in  Sect.~\ref{sec:cutoff_and_ss}. Finally, the RSF model is fully defined once a dynamical evolution equation for $\phi$, which connects the no-sliding $\phi\=t$ and sliding $\phi\=D/|v|$ fixed-points, is provided. This is commonly done through the popular ``aging law''~\cite{Dieterich1986, Marone1998a}
\begin{equation}
\dot \phi =1- \frac{\phi|v|}{D} \ ,
\label{eq:phidot}
\end{equation}
which together with Eq.~\eqref{eq:RSF}, constitute the conventional RSF model. Other $\phi$ evolution equations, $\dot\phi\=\Phi(\phi|v|/D)$ with other functions $\Phi(\cdot)$, are considered and studied in the literature~\cite{Ruina1983,Chester1992,Perrin1995,Putelat2015}, but are not discussed here.

\subsection{Limitations of conventional RSF models and proposed modifications}
\label{sec:limitations_and_modification}

RSF models have been quite extensively used in macroscopic modeling of frictional phenomena.
Yet, we believe that while these models are quite successful in capturing many physical aspects of frictional phenomenology, they are also deficient in various important ways.
Our goal here is to highlight these limitations and to propose modifications to be incorporated into an extended friction model.

\subsubsection{Linear reversible response and an extended RSF model}
\label{sec:linearResponse}

Conventional RSF models predict the frictional stress, cf.~Eq.~\eqref{eq:RSFsinh}, which should provide physically meaningful information both in the presence of sliding ($v\!\ne\!0$) and in its absence ($v\!=\!0$). Using conventional terminology, it should describe both static and dynamic friction. However, Eq.~\eqref{eq:RSFsinh} predicts $\tau(\phi,v\!=\!0)\=0$ in the limit $v\!\to\!0$, i.e.~it predicts vanishing static friction. This happens because $\tau_s(v)$ in Eq.~\eqref{eq:tausinh} corresponds to a Newtonian fluid behavior at small sliding velocities, i.e.~$\tau_s(v\!\to\!0)\!\propto\!v$. One way in which this problem is circumvented in the literature is to change the $v$-dependent term in Eq.~\eqref{eq:RSF} to be $\alpha\log(1+v/v_c)$, such that it vanishes for $v\=0$, together with interpreting the remaining part, $f_0+\beta\log(t/\phi^*)$ (note that $\phi\=t$ for $v\=0$), as an \textit{upper threshold}. That is, in this case $f(\phi, v\=0)$ does not correspond to the frictional stress, but rather to the threshold for the onset of sliding. Mathematically, it means that $f(\phi, v\=0)$ is used to define an inequality imposed on a frictional stress that is calculated from other considerations (in fact, from the bulk shear stress $\sigma_{xy}$ of Eq.~\eqref{eq:elastodynamics}). At the same time, $f(\phi,v)$ of the very same Eq.~\eqref{eq:RSF}, is interpreted as the actual frictional stress for $v\!\ne\!0$, implying that $f(\phi,v)$ has a different physical meaning for $v\=0$ and for $v\!\ne\!0$.

Another common way to circumvent the vanishing static friction problem is by forcing the frictional interface to slide with a small background velocity $v\!\ne\!0$ and use Eq.~\eqref{eq:RSF} as is~\cite{Perrin1995, Zheng1998}. While these ad hoc solutions to the stated problem might provide a practical way out under some circumstances, from a basic physics perspective we expect $\tau(\phi,v)$, or $f(\phi,v)$, to continuously describe the frictional resistance stress under both static and dynamic conditions. In particular, it should properly account for the transition between the two behaviors, e.g.~for the onset of sliding and for sliding direction reversal (which is relevant for cyclic frictional motion, appearing in many engineering systems).

\begin{figure}[ht]
        \centering
        \includegraphics[width=.9\textwidth]{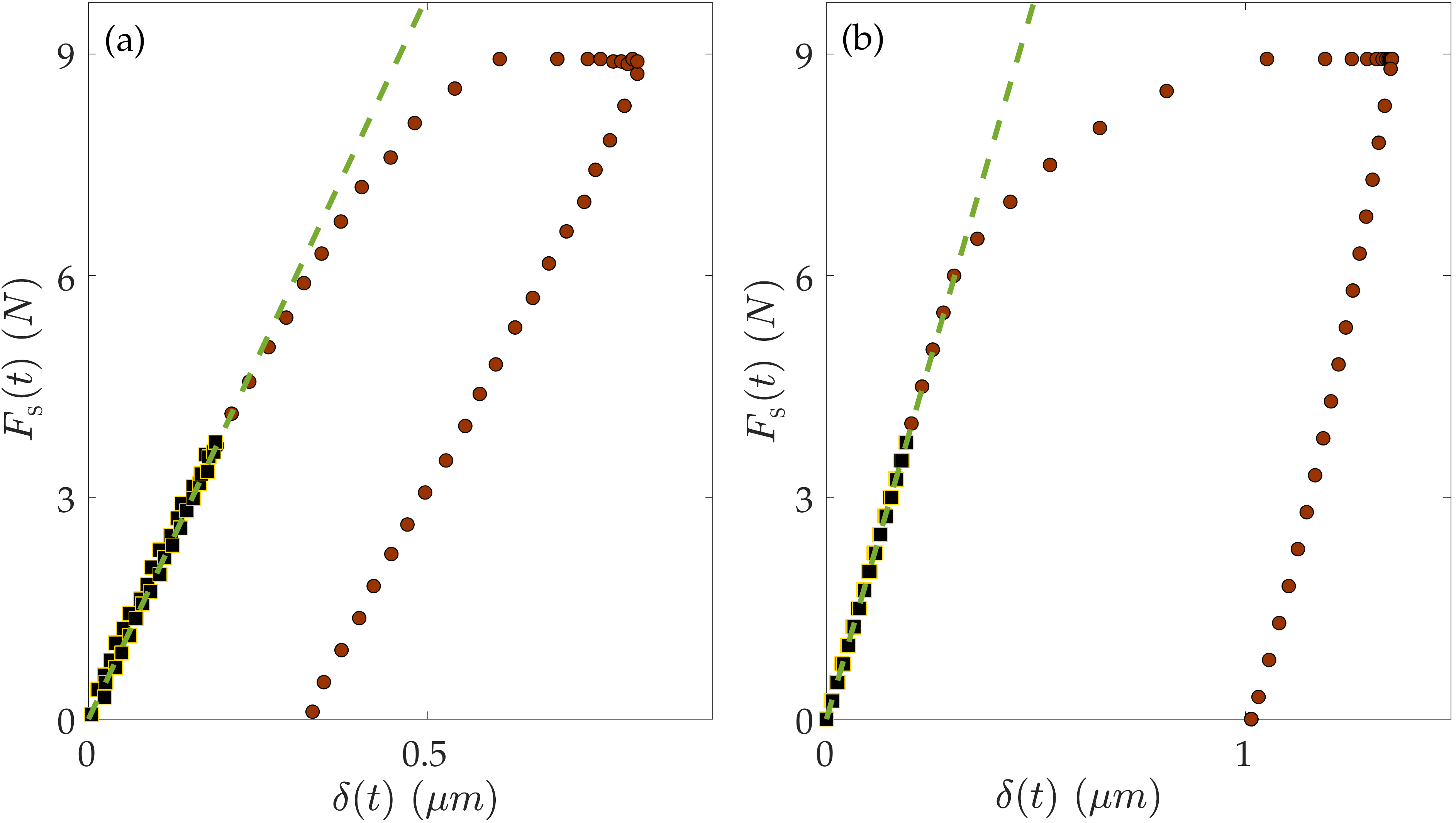}
        \caption{Frictional dynamics under small stresses and the existence of a linear reversible frictional response. (a) Experimental data, taken from Fig.~1 of~\cite{Berthoud1998}. The shear force is shown as a function of slip displacement, with points equally spaced in time. The squares show the loading-unloading protocol at very low stresses, whereas the circles show the load-hold-unload protocol, still below the static friction threshold (see Sec.~\ref{sec:linearResponse}). The former exhibit almost entirely linear reversible response, with essentially no hysteresis. The latter exhibit a transition to irreversible (dissipative) slip, creep dynamics (during the hold phase) that may indicate arrest after a finite time (as suggested by the increasing density of the circles) and a residual slip upon unloading. (b) Our friction model, which includes a linear reversible frictional response, semi-quantitatively reproduces the experimental observations (see text and~\hyperref[sec:interfacial elasticity]{Appendix B} for details).}
        \label{fig:IE}
\end{figure}

In our view, these problems originate from the fact that the conventional RSF model does not feature a \textit{linear reversible response} at low stresses, corresponding to the elastic (reversible) deformation of contact asperities (note that, as explained above, the model does feature a Newtonian viscous behavior, i.e.~a linear {\em irreversible} response). The existence of a reversible response has been experimentally demonstrated~\cite{Courtney-Pratt1957, Archard1957, Berthoud1998, Bureau2000, Popov2010}, see also Fig.~\ref{fig:IE}a here, and is expected on general grounds~\cite{Campana2011}. That is, we expect the asperity-level stress $\tau_s$ to be {\em finite and reversible} under small static shear loading conditions due to the elastic deformation of contact asperities. This should result in $\tau(\phi,v)$ that is valid for both $v\!=\!0$ and $v\!\ne\!0$, as well as for $v$ that changes sign dynamically, which should resolve the problems with conventional RSF models stated above. Indeed, some authors have incorporated an interfacial elastic response in their modeling efforts~\cite{Povirk1993, Bureau2000, Coker2005a, Shi2008, Braun2009, Shi2010, Bouchbinder2011, Rubinstein2011, Bar-Sinai2012, Bar-Sinai2013} and noted its potential importance for various frictional phenomena, such as the onset of sliding~\cite{Popov2010} and rupture mode selection~\cite{Shi2008}. Here we describe an extended RSF-type model that incorporates a linear reversible response at low stresses. The model has been briefly introduced and studied in previous publications~\cite{Bouchbinder2011, Bar-Sinai2012, Bar-Sinai2013}.

The basic idea is to incorporate into the RSF framework an elastic interfacial response at small stresses, within a Kelvin-Voigt-like viscoealstic model~\cite{Ferry1980,Thomson1865} together with threshold dynamics. To that aim, we reinterpret the asperity-level stress in Eq.~\eqref{eq:tausinh} as a viscous-like stress $\tau_s^{vis}(\phi,v)$, which
is a partial stress constituting only one contribution to the total asperity-level stress $\tau_s$. The
complementary contribution is the asperity-level elastic stress $\tau_s^{el}$, such that
$\tau_s\=\tau_s^{vis}+\tau_s^{el}$. While this way of introducing asperity-level elasticity into the
friction model is not unique, it suffices for our present purposes, as will be shown below. Using the Bowden and Tabor relation $\tau\=A(\phi)\tau_s(v)$
to translate the asperity-level stress to a coarse-grained stress, we obtain
\begin{equation}
\tau(\phi,v,\tau^{\rm el}) =\tau^{\rm el} + \tau^{\rm vis}(\phi,v)\ ,
\label{eq:KV}
\end{equation}
where, as explained above, $\tau^{\rm vis}(\phi,v)\=A(\phi)\tau_s(v)$  with $\tau_s(v)$ of Eq.~\eqref{eq:tausinh}.

In Eq.~\eqref{eq:KV}, $\tau^{\rm el}$ is a dynamic variable by itself, in addition to $\phi$ and $v$. Therefore, we need to derive its evolution equation.
As explained above, at small slip displacements we expect the interface to respond linear elastically. To capture this basic physical behavior, we treat the interface as a boundary-layer of height $h$, much smaller than the macroscopic size $H$ of the system, which is characterized by an effective shear modulus $\mu_0$.
This effective modulus may be different from the shear modulus $\mu$ of the bulk due to the existence of stress-free surfaces along the interface (i.e.~surface roughness).
Under deformation, the total shear accumulated in the boundary layer is $\delta/h$, producing a shear stress $\mu_0 \delta/h$. Such an elastic response of the interface has been directly measured~\cite{Courtney-Pratt1957, Berthoud1998, Bureau2003}, as shown in Fig.~\ref{fig:IE}a, and has been predicted theoretically~\cite{Campana2011}. Defining a coarse-grained elastic stress $\tel$ through $A_n\tel\=A_r\mu_0 \delta/h$ and differentiating it with respect to time, we obtain $\dot\tau^{\rm el}\=\mu_0 A\,v/h$ (where we assumed that $A$ varies much slower than $\tel$ such that its time-derivative can be neglected). The term $\mu_0 A\,v/h$ constitutes only one contribution to $\dot\tau^{\rm el}$; the other contribution is discussed next.

Irreversible sliding drives contact annihilation and renewal, and hence is accompanied by a reduction in the average elastic stress stored in the contact asperities.
For simplicity, we assume a linear form for this reduction, which is similar to that of $\phi$ in Eq.~\eqref{eq:phidot}. Thus, the evolution equation of $\tau^{\rm el}$ is taken to be
\begin{equation}
 \dot\tau^{\rm el}= \frac{\mu_0 }{h}A\,v-\tel \frac{|v|}{D}g(\tau,v)\ ,
 \label{eq:tauel}
\end{equation}
where the function $g(\tau,v)$ controls the transition between the reversible and irreversible regimes. In fact, the very same function should appear in a modified version of Eq.~\eqref{eq:phidot}, which takes the form $\dot \phi\=1-\phi|v|g(\tau,v)/D$ and is adopted here. When $g\!\ll\!1$, the system responds elastically, i.e.~$\tau^{\rm el}\!=\!\mu_0 A \delta/h$. When $g(\cdot)\!\sim\!{\cal O}(1)$, irreversible stress and area reduction take place. Therefore, $g(\cdot)$ plays the role of an effective threshold for the onset of irreversible slip, as in conventional elasto-plastic models that feature a yield stress. Explicit expressions for $g(\cdot)$ will be discussed below in relation to various applications.

To directly demonstrate the physical relevance and importance of interfacial elasticity, we focus on the experimental results of~\cite{Berthoud1998}, which are reproduced here in Fig.~\ref{fig:IE}a. In these experiments, the multi-contact interface between two PMMA blocks has been subjected to two time-dependent shear loading-unloading protocols below the 'static' threshold for gross slip (which for this experimental system corresponded to a shear force of $F_{\rm static}\=10$N). In the first protocol (the squares in Fig.~\ref{fig:IE}a, equally spaced in time), the shear force $F_{_{\rm S}}(t)$ is externally increased at a constant (and small) rate up to $4$N (i.e.~about $40\%$ of the nominal 'static' friction threshold) and then decreased back to zero at the same small rate. The observed response, i.e.~the slip displacement $\delta(t)$, is nearly linear in the driving force $F_{_{\rm S}}(t)$, featuring very little dissipation (i.e.~hysteresis) and almost full recovery of the initial state. These observations explicitly demonstrate the existence of a linear reversible (elastic) response. In the second protocol (the circles in Fig.~\ref{fig:IE}a, equally spaced in time), the shear force $F_{_{\rm S}}(t)$ is externally increased at a constant (and small) rate up to $9$N (i.e.~about $90\%$ of the nominal 'static' friction threshold), it is then maintained at this level for sufficiently long time and eventually it is decreased back to zero at the same small rate. Under this load-hold-unload protocol, the slip displacement $\delta(t)$ is no longer linear in the driving force $F_{_{\rm S}}(t)$ for values larger than $\simeq\!7$N. Moreover, when $F_{_{\rm S}}(t)$ is kept at its maximal value (here of $9$N), $\delta(t)$ exhibits creep dynamics that may indicate arrest after a finite time (as suggested by the increasing density of the circles). Finally, as the shear force is reduced back to zero, a finite residual (irreversible) slip remains. This behavior corresponds to a viscoealstic response with threshold dynamics.

The friction model discussed above, which incorporates an interfacial elastic response at small driving forces, naturally and semi-quantitatively reproduces these experimental observations. To see this, we neglect all bulk degrees of freedom, whose role has been deliberately minimized in the experiments of~\cite{Berthoud1998} in order to isolate the interfacial response (this is the only place in this paper where bulk degrees of freedom are neglected), and focus on the equation
\begin{equation}
\label{eq:1DOF_F}
F_{_{\rm S}}(t) = F_{_{\rm N}} f(\phi,\dot\delta,\tau^{\rm el}) = A_n\tau(\phi,\dot\delta,\tau^{\rm el})\ .
\end{equation}
Here $F_{_{\rm S}}(t)$ is the applied shear force (i.e.~corresponding to each one of the two protocols discussed above), $F_{_{\rm N}}$ is the constant applied normal force and $A_n$ is the nominal (macroscopic) contact area (such that $\sigma_0\=F_{_{\rm N}}/A_n$ is the normal stress, not used here). $\tau(\phi,\dot\delta,\tau^{\rm el})$ of Eq.~\eqref{eq:1DOF_F} is given in Eq.~\eqref{eq:KV}, $\tau^{\rm el}$ evolves according to Eq.~\eqref{eq:tauel} and $\phi$ according to $\dot \phi\=1\!-\!\phi|v|g(\tau,v)/D$ (the choice of $g(\cdot)$ is discussed in~\hyperref[sec:interfacial elasticity]{Appendix B}). The solutions of these equations for the two experimental protocols discussed in relation to Fig.~\ref{fig:IE}a are presented in Fig.~\ref{fig:IE}b (see additional details in~\hyperref[sec:interfacial elasticity]{Appendix B}), reproducing all of the experimental observations semi-quantitatively (no attempt has been made to quantitatively reproduce the experimental data). This example concludes the discussion of the first limitation we identify with RSF models, i.e.~the absence of a linear reversible response at small stresses.

\subsubsection{Short time cutoff and the steady-state friction curve}
\label{sec:cutoff_and_ss}

The second problem we identify with conventional RSF models is related to Eq.~\eqref{eq:Awrong}. Imagine that a fresh interface is formed by bringing two bodies into frictional contact. In the absence of irreversible slip, for which $\dot \phi\=1$, the age $\phi$ can be replaced by the time $t$ elapsed from the formation of the interface. Equation \eqref{eq:Awrong} then predicts that the contact area will grow logarithmically with time, as observed in a broad range of experiments~\cite{Kilgore1993, Nakatani2006, Ben-David2010, Nagata2012}. However, when $t\!\to\!0$ it becomes singular, which is clearly unphysical. Evidently, a short time cutoff exists and must be incorporated. A simple way to do so is to replace Eq.~\eqref{eq:Awrong} by
\begin{equation}
 A(\phi)=\frac{\sigma}{\sigma_{\hbox{\tiny H}}}\left[1+b\log\left(1+\frac{\phi}{\phi^*}\right)\right]\ .
 \label{eq:A}
\end{equation}
With this modification, $\phi^*$ is no longer an arbitrary time scale --- it is an intrinsic interfacial/material parameter which can be interpreted as a typical cutoff time for the onset of logarithmic aging. The existence of this cutoff has been directly verified experimentally~\cite{Marone1998, Nakatani2006, Ben-David2010}; it has been suggested by Dieterich \cite{Dieterich1978, Dieterich1979}, mentioned by Baumberger, Caroli and coworkers \cite{Baumberger1999, Bureau2001, Baumberger2006} and extensively discussed recently in~\cite{Bar-Sinai2012, Bar-Sinai2014}.

While this modification might appear at first glance rather innocent, it has major implications for persistent sliding because the short time cutoff is translated, through the steady-state relation $\phi\=D/|v|$, to a high-velocity cutoff. To see this, note that Eq.~\eqref{eq:A} predicts that $A$ decreases logarithmically with the steady sliding velocity $v$. The presence of a short time cutoff implies that this decrease saturates at velocities $\gtrsim\!D/\phi^*$. At higher velocities the real contact area approaches a constant, and the velocity dependence of friction is dominated by $\tau_s(v)$. Since the latter is intrinsically an increasing function of $v$ (cf. Eq.~\eqref{eq:tausinh}), friction becomes \textit{velocity-strengthening} in this range of velocities. Put differently, the short time cutoff makes the steady sliding frictional resistance a non-monotonic function of the sliding velocity, reaching a minimum at $v_{\rm min}\!\simeq\!D/\phi^*$. As will be thoroughly discussed below, the crossover from velocity weakening to velocity-strengthening has important implications on many aspects of the spatiotemporal dynamics of frictional systems.
\begin{figure}[ht]
  \centering
  \includegraphics[width=.75\textwidth]{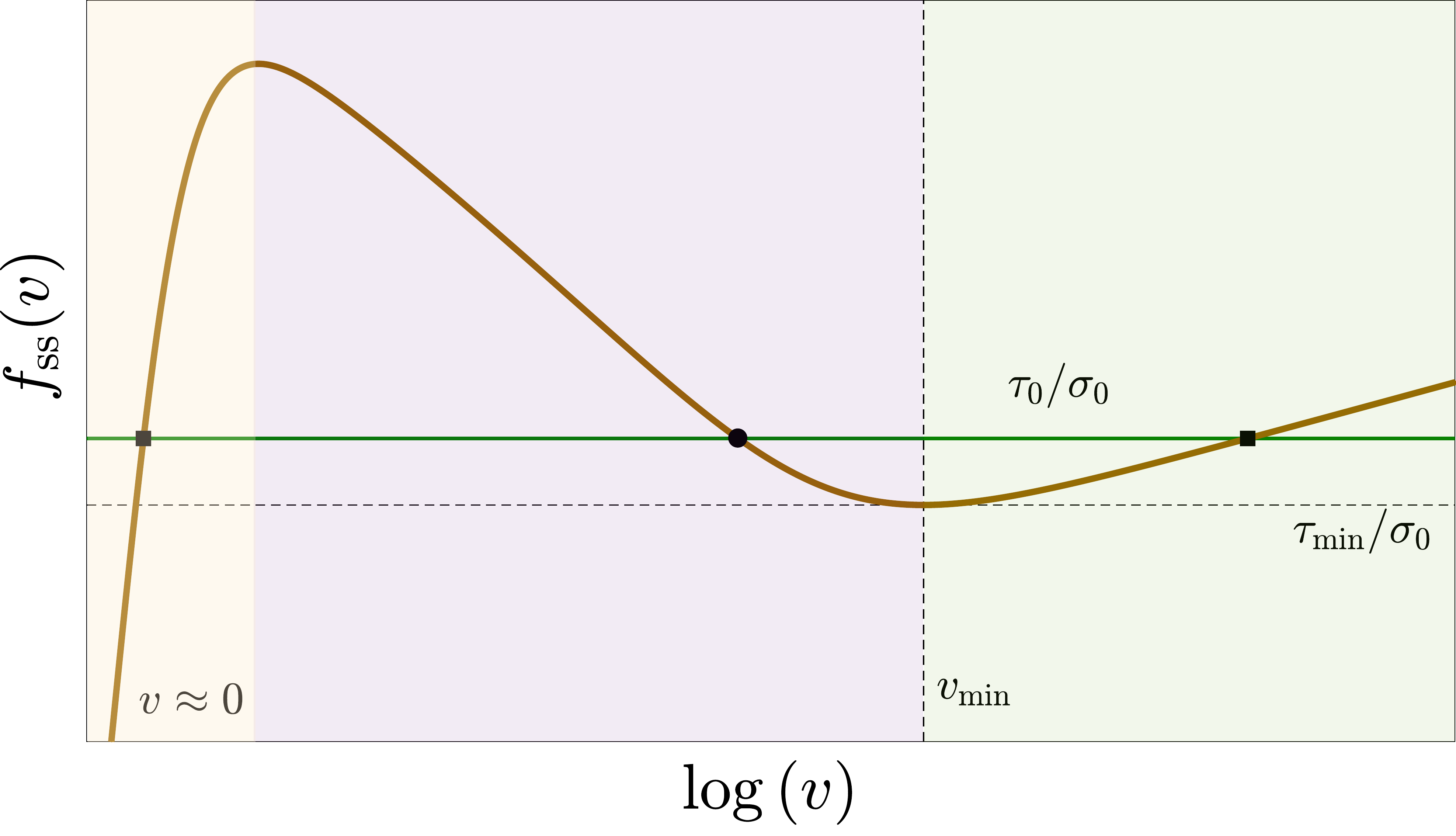}
  \caption{The generic form of steady-state frictional resistance $f_{\rm ss}(v)$ (see text and~\hyperref[sec:math_friction]{Appendix A} for details). At extremely low sliding velocities (yellow-shaded), steady-state friction is velocity-strengthening. This regime, associated with $v\!\approx\!0$ in the figure, corresponds to stick conditions (effectively no-sliding) either because the slip velocity is minute or because steady-state is dynamically inaccessible due to extremely long relaxation timescales (see text for additional discussion). At intermediate velocities (purple-shaded), friction is velocity-weakening. Finally, at higher velocities (green-shaded), above a local minimum occurring at $v_{\rm min}$ (corresponding to a frictional resistance $\tau_{\rm min}/\sigma_0$), friction becomes velocity-strengthening again. When loaded at stresses higher than this minimal value, $\tau_0\!>\!\tau_{\rm min}$ (horizontal line), three homogeneous solutions (fixed-points) exist; two stable (on the velocity-strengthening branches, marked by squares) and one unstable (on the velocity-weakening branch, marked by a circle), see text for additional discussion.}
  \label{fig:ss_friction}
\end{figure}

We conclude this discussion of conventional RSF models and their extensions (the complete set of equations defining the constitutive law is collected in~\hyperref[sec:math_friction]{Appendix A}) by stating what we perceive to be a generic picture of the steady-sliding friction curve. The latter, denoted by $f_{\rm ss}(v)$ and derived in~\hyperref[sec:math_friction]{Appendix A}, is schematically plotted in Fig.~\ref{fig:ss_friction}. The function $f_{\rm ss}(v)$ has a generic ``N-like'' shape, composed of three distinguishable regimes. First, at very small slip velocities, the frictional stress increases linearly with $v$, corresponding to a viscous-like creep motion~\cite{Estrin1996}. We note that at these low velocities, dynamically reaching the true steady-state might be very difficult/long because the relaxation time scales as $D/|v|$ (though there exist experimental measurements of this regime~\cite{Shimamoto1986}). This velocity-strengthening branch (i.e.~an increase in the frictional resistance with increasing steady-state sliding velocity) corresponds in practical terms to `stick conditions', i.e.~to situations in which the frictional interface is essentially locked-in; the reason for this is either the extremely low slip velocities in this regime or that it takes so long to reach this low velocity steady-state that the interfacial response is in fact predominantly elastic (representing an extremely long transient). This regime is terminated by a maximum of the curve, followed by a logarithmic velocity-weakening regime (corresponding to the $\alpha\!<\!\beta$ case discussed in Sect.~\ref{sec:RSF} in relation to Eq.~\eqref{eq:RSF}). Finally, the curve reaches a minimum around $v_{\rm min}\!\simeq\!D/\phi^*$ and then increases logarithmically (i.e.~follows a velocity-strengthening behavior, as discussed above).

The validity of the N-shaped steady-state friction curve of Fig.~\ref{fig:ss_friction} is supported by experimental data for a broad range of materials~\cite{Bar-Sinai2014} and its implications for the dynamics of frictional systems will be reviewed below. We note that while we consider here logarithmic velocity-strengthening above the minimum of the steady-state curve, strengthening can in fact become stronger than logarithmic, as discussed in~\cite{Baumberger2006,Bar-Sinai2014,Bar-Sinai2015a}. We also note that at yet higher slip velocities (compared to those shown in Fig.~\ref{fig:ss_friction}), the heat generated by frictional dissipation might not have enough time to diffuse away from the interface, leading to thermal softening accompanied by a substantial drop in the frictional resistance~\cite{DiToro2004, Goldsby2011}. This regime is not shown in Fig.~\ref{fig:ss_friction} and is not further discussed in this paper. Finally, note that the velocity-weakening and post-minimum velocity-strengthening regimes of the steady-state friction curve are still approximately described by Eq.~\eqref{eq:RSF} once $\beta\log(\phi/\phi^*)$ is replaced by $\beta\log(1+\phi/\phi^*)$ (following Eq.~\eqref{eq:A}), but with redefined $f_0$, $\alpha$ and $\beta$ (see~\hyperref[sec:math_friction]{Appendix A} for details). For $\alpha\!>\!\beta$, the steady-state friction curve is velocity-strengthening for all slip velocities (again, if extremely high-velocity thermal softening is excluded from the discussion). With this, we conclude our discussion of the generic properties of interfacial constitutive laws.

\section{Bulk elasticity in frictional systems}
\label{sec:bulk_elasticity}

In the previous section we focused on the generic properties of the interfacial constitutive law, i.e.~on $f\!\left[\delta(x,t), \partial_t\delta(x,t), ...\right]$ in Eq.~\eqref{eq:coupling}. Our goal here is to discuss the properties of the bulk-mediated interaction contributions to this basic equation, i.e.~${\cal F}_{_{\!\tau}}\!\left[\partial_x\delta(x,t), \partial_t\delta(x,t)\right]$ and ${\cal F}_{_{\!\sigma}}\!\left[\partial_x\delta(x,t),\partial_t\delta(x,t)\right]$. The latter, ${\cal F}_{_{\!\sigma}}\!\left[\partial_x\delta(x,t),\partial_t\delta(x,t)\right]$, represents the coupling between interfacial slip and normal stress variations along the interface. It is now well-established that this coupling exists only when reflection symmetry with respect to the frictional interface is broken~\cite{Rice2001,Weertman1980, Aldam2016,Heimisson2019}. This symmetry can be broken by material contrast (i.e.~having two blocks made of different materials, cf.~Fig.~\ref{fig:sketch}), by geometric asymmetry (i.e.~having two blocks of identical materials, but with different shapes, cf.~Fig.~\ref{fig:sketch}) or by the loading configuration. In the first part of the discussion in this section, we assume perfect material and geometric symmetry, and perfectly anti-symmetric loading, such that ${\cal F}_{_{\!\sigma}}\!\left[\partial_x\delta(x,t),\partial_t\delta(x,t)\right]\=0$. Consequently, we first focus on ${\cal F}_{_{\!\tau}}\!\left[\partial_x\delta(x,t), \partial_t\delta(x,t)\right]$.

The shear stress contribution ${\cal F}_{_{\!\tau}}\!\left[\partial_x\delta(x,t), \partial_t\delta(x,t)\right]$ generally features complex spatiotemporal properties. The spatial properties, in particular the range of bulk-mediated interaction between different parts of the interface, is determined by the system dimensions. We assume that the system length $L$ is much larger than its height $H$ (cf.~Fig.~\ref{fig:sketch} for a visual definition of $L$ and $H$), and focus on the effect of the latter on ${\cal F}_{_{\!\tau}}\!\left[\partial_x\delta(x,t), \partial_t\delta(x,t)\right]$. Consider first two infinite blocks, i.e.~$H\!\to\!\infty$, for which we have~\cite{Geubelle1995,Morrissey1997,Breitenfeld1998}
\begin{equation}
{\cal F}_{_{\!\tau}}\!\left[\partial_x\delta(x,t), \partial_t\delta(x,t)\right]= \frac{\mu}{2c_s}\partial_t\delta(x,t) - s(x,t) \ ,
\label{eq:infiniteH_dynamic}
\end{equation}
where $\mu$ is the shear modulus (already defined above) and $c_s$ is the shear wave-speed ($s(x,t)$ will be defined and briefly discussed below). The right-hand-side of Eq.~\eqref{eq:infiniteH_dynamic} represents the infinite half-space linear elastodynamic Green's function~\cite{Geubelle1995,Morrissey1997,Breitenfeld1998}, corresponding to the momentum balance equation $\rho \ddot{\bm u}\=\nabla\!\cdot\!{\bm \sigma}$ (where $\rho$ is the mass density and the stress tensor $\bm\sigma$ is related to displacement gradients $\nabla{\bm u}$ through Hooke's law), and contains two contributions. The contribution $\mu\partial_t\delta(x,t)/2c_s$ is `instantaneous', i.e.~it depends on the slip velocity $\partial_t\delta(x,t)$ alone, and corresponds to wave radiation from the interface into the bulks forming it~\cite{Ben-Zion1995,Perrin1995,Zheng1998,Crupi2013}. It is known as the `radiation-damping' term because mathematically it appears as an effective linear drag/viscous stress that the bulks apply on the interface.

The contribution $s(x,t)$ in Eq.~\eqref{eq:infiniteH_dynamic} is infinitely long-ranged in space, i.e.~it involves integration over the whole interface with power-law kernels, and involves time integration that is limited by causality (formally it is a functional of $\delta(x,t)$, not just a function of $x$ and $t$). In general, $s(x,t)$ does not admit an analytic real-space representation and even its spectral (Fourier) representation may be highly complex~\cite{Geubelle1995,Morrissey1997,Breitenfeld1998}. In order to nevertheless gain some physical intuition into $s(x,t)$ and to support the statements just made about it, we consider situations in which the frictional dynamics are significantly slower compared to elastodynamic effects. That is, we consider the quasi-static limit in which the inertia of the bulks plays no role. In this quasi-static limit the radiation-damping term is also negligibly small and we have (see, for example,~\cite{Weertman1965})
\begin{equation}
{\cal F}_{_{\!\tau}}\!\left[\partial_x\delta(x,t), \partial_t\delta(x,t)\right]\simeq-\frac{\mu}{2\pi}\int^\infty_{-\infty}\frac{\partial_{x'}\delta(x',t)}{x'-x}dx' \ .
\label{eq:QS}
\end{equation}
The infinitely long-range bulk-mediated interaction between different parts of the interfaces and the associated power-law kernel are evident, as is also schematically illustrated in Fig.~\ref{fig:sketch}. Mathematically handling the infinitely long-range interaction in the $H\!\to\!\infty$ limit generally raises significant difficulties and in many case a variety of numerical methods are invoked, e.g.~spectral techniques~\cite{Geubelle1995,Morrissey1997,Breitenfeld1998}.

The spatiotemporal range of interaction encapsulated in ${\cal F}_{_{\!\tau}}\!\left[\partial_x\delta(x,t), \partial_t\delta(x,t)\right]$ diminishes as the system height $H$ is reduced. A particularly useful and analytically tractable expression is obtained in the limit of small $H$. That is, if one assumes that $H$ is sufficiently smaller than any other relevant lengthscale in the problem, a systematic expansion (reminiscent of the commonly used shallow water approximation in fluid mechanics~\cite{Acheson1990}) yields~\cite{Bar-Sinai2012,Bar-Sinai2013,Viesca2016b}
\begin{equation}
{\cal F}_{_{\!\tau}}\!\left[\partial_x\delta(x,t), \partial_t\delta(x,t)\right]\simeq\rho H \pa_{tt}\delta(x,t) - \bar{\mu} H\pa_{xx}\delta(x,t) \ .
\label{eq:1d}
\end{equation}
Here $\bar\mu$ is proportional to $\mu$, with a prefactor that depends on Poisson's ratio~\cite{Bar-Sinai2013}. In this thin-systems approximation, sometimes termed the quasi-1D limit, the spatiotemporal interaction becomes short-ranged (i.e.~captured by second order differential operators, with no integral terms) and the bulk --- described by a scalar wave equation with $H$-dependent coefficients --- is effectively inseparable from the interface. Precisely for this reason, the radiation-damping term of Eq.~\eqref{eq:infiniteH_dynamic} also disappears (i.e.~there is no distinct bulk to radiate energy into) and ${\cal F}_{_{\!\sigma}}\=0$. While the short-range spatiotemporal interaction is a limitation, the relative simplicity of the small-$H$ formulation allows a comprehensive, quantitative and predictive analytical treatment of complex spatiotemporal frictional dynamics, in addition to extremely efficient numerical calculations~\cite{Bar-Sinai2012,Bar-Sinai2013,Bar-Sinai2015a}, as will be demonstrated below.

When $H$ is neither infinitely large nor small, interesting new effects related to the finite system height emerge, as will be discussed below in relation to various physical phenomena. Moreover, when the perfect symmetry assumption is relaxed, and reflection symmetry with respect to the frictional interface is broken by material contrast, geometric asymmetry or the loading configuration, ${\cal F}_{_{\!\sigma}}\!\left[\partial_x\delta(x,t),\partial_t\delta(x,t)\right]$ does not vanish. In such situations, which are quite prevalent, one needs to consider both ${\cal F}_{_{\!\tau}}\!\left[\partial_x\delta(x,t), \partial_t\delta(x,t)\right]$ and ${\cal F}_{_{\!\sigma}}\!\left[\partial_x\delta(x,t),\partial_t\delta(x,t)\right]$, and rich dynamical effects emerge from Eq.~\eqref{eq:coupling}. While the analysis of finite systems can be pursued analytically in some cases, numerical methods such as the Finite Element Method (FEM) are required in many other cases. Finally, we note that finite-$H$ systems, i.e.~realistic systems, are well described by the infinite-$H$ formulation discussed above on time scales shorter than the typical wave reflection time from outer boundaries (which is determined by $H/c_s$). In the next section we will see how the various forms of bulk elasticity discussed in this section couple to the interfacial constitutive law to give rise to rich spatiotemporal frictional dynamics.

\section{Basic phenomena in spatially-extended frictional systems}
\label{sec:phenomena}

Our goal in this section is to demonstrate, through a broad range of examples, how basic phenomena in spatially-extended frictional systems emerge from the inherent coupling between the interfacial constitutive law (discussed in Sect.~\ref{sec:constitutive}) and bulk elasticity (discussed in Sect.~\ref{sec:bulk_elasticity}).

\subsection{The stability of homogeneous sliding}
\label{subsec:LSA}

A central question in relation to spatially-extended frictional systems concerns the stability of homogeneous sliding~\cite{Rice1983,Rice2001}. That is, suppose that two bodies in frictional contact slide at a constant homogeneous slip velocity $v_0$. When this sliding motion is unstable, interesting phenomena such as squeaky door hinges~\cite{Thomsen1999}, squealing car brakes~\cite{Rhee1991} or earthquakes along geological faults~\cite{Brace1966,Dieterich1975} might emerge. What are the conditions for instability? How are they influenced by the interplay between interfacial friction and bulk elasticity? In this section, we scratch the surface of these important questions through two examples that highlight the interplay between interfacial friction and bulk elasticity in relation to the stability of frictional systems. Mathematically, one considers small spatiotemporal perturbations $\delta{v}(x,t)$ on top of the homogeneous sliding state $v_0$, i.e.~$v(x,t)\=v_0+\delta{v}(x,t)$, and expand all relevant equations to leading (linear) order in $\delta{v}(x,t)$. Due to linearization, one can consider each Fourier mode separately, i.e.~express the perturbation as $\delta{v}(x,t)\!\sim\!\exp(\Lambda t - ikx)$, where $\Lambda$ is the complex frequency and $k$ is the wavenumber. An instability emerges whenever the real part of $\Lambda$ becomes positive, i.e.~$\Re[\Lambda]\!>\!0$, which implies that perturbations exponentially grow in time.

The complex frequency $\Lambda$ and the wavenumber $k$ can be related by perturbing Eq.~\eqref{eq:coupling} to linear order in $\delta{v}(x,t)$. The result takes the schematic form
\begin{equation}
-\delta\!{\cal F}_{_{\!\tau}} = \sigma_0\,\delta{\!f} - f\,\delta\!{\cal F}_{_{\!\sigma}} \ ,
\label{eq:perturbation}
\end{equation}
where $\delta\{\cdot\}$ (not to be confused with the slip displacement $\delta(x,t)$) represents a variation with respect to $v$.

\begin{figure}[ht]
  \centering
  \includegraphics[width=0.9\textwidth]{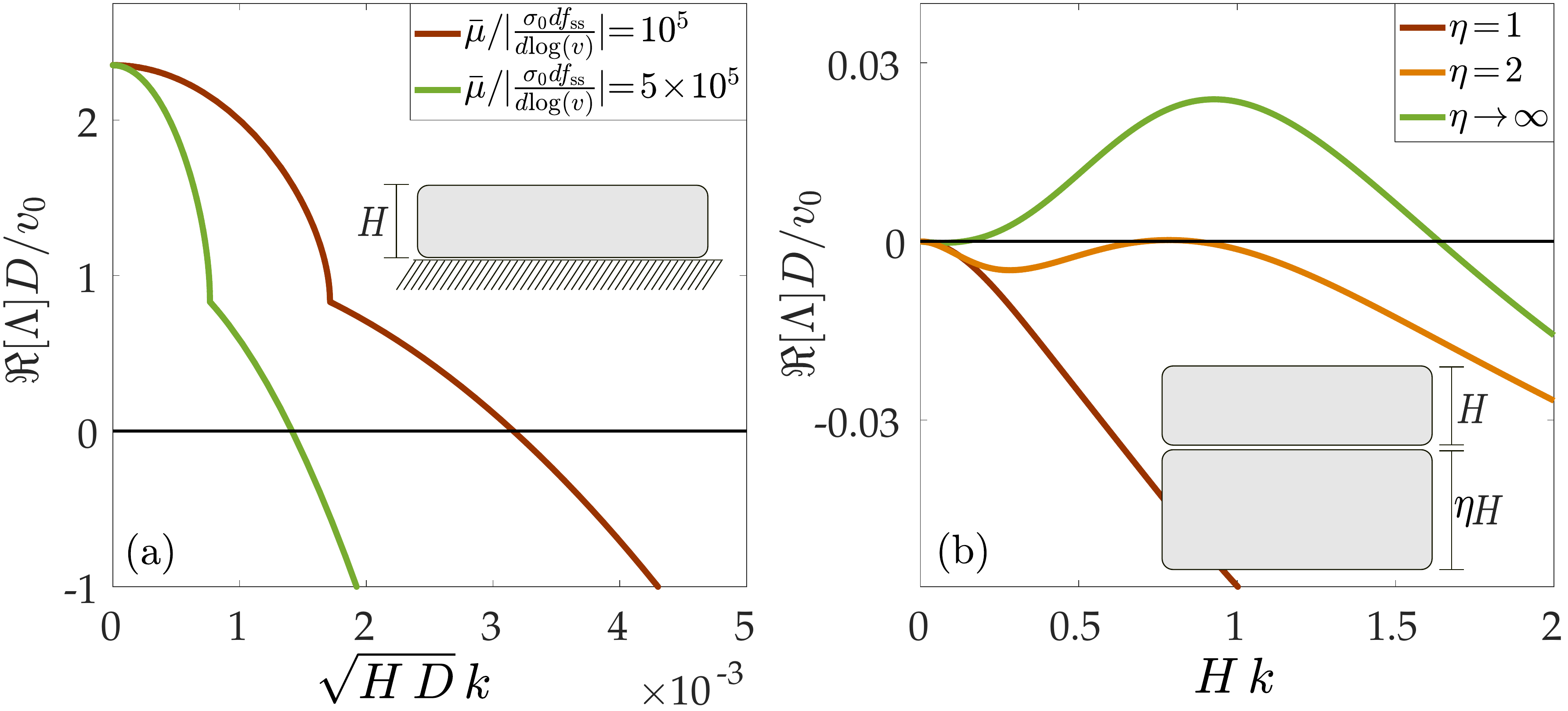}
  \caption{Linear stability spectrum of homogeneous steady sliding. Plotted is the growth rate of perturbations $\Re[\Lambda]$ as a function of wavenumber $k$ (both properly non-dimensionalized). When $\Re[\Lambda]\!>\!0$, sliding is unstable. (a) In the small-$H$ limit, sliding on top of a rigid substrate becomes unstable when $k$ is smaller than a critical value, determined by the critical nucleation length $L_c$ of Eq.~\eqref{eq:Lc}. Curves are shown for two parameter sets, demonstrating that stability is promoted (i.e.~the range of unstable modes shrinks) when the body is stiffer (large $\bar\mu$) and when friction is less velocity-weakening (smaller $\sigma_0df_{\rm ss}/d\log(v)$). Details of the derivation of the linear stability spectrum can be found in the Supporting Information of~\cite{Aldam2017a}, though similar calculations appeared earlier~\cite{Rice1983,Rice2001}. Note that the kink in the spectrum is related to its complex-plane structure. (b) For a system of two elastic bodies of different height (quantified by the height contrast $\eta$), velocity-strengthening sliding can be destabilized by geometric contrast/asymmetry alone (adapted from Fig.~5 of~\cite{Aldam2016}). In the absence of geometric contrast ($\eta\!=\!1$), sliding is categorically stable. At a critical contrast ($\eta_c\!\approx\!2$), an instability emerges (first zero crossing). Finally, for large contrast ($\eta\!\gg\!1$), a range of unstable modes with $k\!\sim\!H^{-1}$ exists (see text for additional details). The insets show the geometric configuration in each panel.}
  \label{fig:LSA}
\end{figure}

The first example we consider in this context concerns the stability of homogeneous sliding in the small $H$ limit in the quasi-static regime, i.e.~when Eq.~\eqref{eq:1d} reads ${\cal F}_{_{\!\tau}}\!\simeq\!-\bar{\mu} H\pa_{xx}\delta(x,t)$. In the small $H$ limit there exists no coupling between slip and normal stress variations, hence ${\cal F}_{_{\!\sigma}}\=0$ in Eq.~\eqref{eq:perturbation}. Finally, $v_0$ is assumed here to be on the velocity-weakening branch of the steady-state friction curve in Fig.~\ref{fig:ss_friction}, i.e.~$\tfrac{df_{\rm ss}(v)}{d\log{v}}\!<\!0$ at $v_0$. In this case, $\delta{\!f}$ in Eq.~\eqref{eq:perturbation} is destabilizing, i.e.~from the perspective of the friction law alone, perturbations are expected to be amplified. However, as will be shown next, the bulk contribution $\delta\!{\cal F}_{_{\!\tau}}$ is stabilizing and consequently it generates a stability regime that depends on an emerging elasto-frictional lengthscale. Performing the analysis~\cite{Rice1983,Aldam2017a}, one finds that homogeneous sliding under the stated conditions is unstable for $0\!<\!k\!<\!k_c$, where $k_c\=2\pi/L_c$ is determined by a critical length of the form~\cite{Bar-Sinai2013}
\begin{equation}
L_c \sim \sqrt{\frac{\bar{\mu}\,H\,D}{-\sigma_0\frac{df_{\rm ss}(v)}{d\log{v}}}} \ .
\label{eq:Lc}
\end{equation}
This result shows that indeed stability is enhanced (i.e.~the range of instability shrinks) as the bulk combination $\bar{\mu}H$ increases compared to the interfacial friction combination $\frac{df_{\rm ss}(v)}{d\log{v}}/D$, and that the competition between the two (along with the applied normal stress $\sigma_0$) gives rise to a characteristic lengthscale. These properties are demonstrated in Fig.~\ref{fig:LSA}a. Finally, note that a similar physical picture emerges in the $H\!\to\!\infty$ limit, where the critical length of Eq.~\eqref{eq:Lc} is replaced by an $H$-independent length of the form $L_c\!\sim\!\frac{\mu\,D}{-\sigma_0\frac{df_{\rm ss}(v)}{d\log{v}}}$~\cite{Dieterich1992}. Results for intermediate $H$ values will be presented in Sect.~\ref{subsec:creep_patches}.

The second example we consider concerns the stability of homogeneous sliding of two finite height bodies made of the same material, where fully inertial dynamics are taken into account (unlike the quasi-static dynamics discussed in the previous example). This time, the homogeneous slip velocity $v_0$ is assumed to be on the velocity-strengthening branch of the steady-state friction curve in Fig.~\ref{fig:ss_friction}, i.e.~$\tfrac{df_{\rm ss}(v)}{d\log{v}}\!>\!0$ at $v_0$. In this case, $\delta{\!f}$ in Eq.~\eqref{eq:perturbation} is stabilizing, i.e.~from the perspective of the friction law alone perturbations are expected to be attenuated, and $\delta\!{\cal F}_{_{\!\tau}}$ is still stabilizing. Consequently, any possible instability can arise only from a destabilizing normal stress contribution $\delta\!{\cal F}_{_{\!\sigma}}$. Let us denote the height of the upper body by $H$ and that of the lower one by $\eta H$~\cite{Aldam2016}. For $\eta\=1$, i.e.~for a perfectly symmetric system, we have $\delta\!{\cal F}_{_{\!\sigma}}\=0$, and homogeneous sliding under the stated conditions is categorically stable~\cite{Dieterich1992}. This is demonstrated in Fig.~\ref{fig:LSA}b. Consider then the case in which $\eta\!\gg\!1$, i.e.~there exists significant geometric asymmetry in the system, giving rise to a destabilizing normal stress contribution $\delta\!{\cal F}_{_{\!\sigma}}$ that competes with the velocity-strengthening stabilizing friction contribution. Depending on the values of the problem's parameters the former can even overcome the latter, implying an instability. This scenario is demonstrated in Fig.~\ref{fig:LSA}b, where an instability emerges at a critical geometric asymmetry level $\eta_c\!>\!1$ and a finite range of unstable modes with wavenumbers $k\!\sim\!H^{-1}$ exists for larger geometric asymmetry levels, showing that bulk geometry and elasticity can destabilize frictional sliding that is otherwise stable~\cite{Rice2001,Aldam2016}.

There are other physical ways in which the interplay between interfacial friction and bulk elasticity affects the stability of frictional systems. For example, the combined effect of bimaterial contrast, finite geometry and inertial dynamics, which strongly affect $\delta\!{\cal F}_{_{\!\sigma}}$, has been shown to play crucial roles in the development of instabilities of velocity-strengthening interfaces (i.e.~when the frictional contribution $\delta{\!f}$ is stabilizing~\cite{Brener2016}). In particular, it has been shown that there exists a universal instability with a wavenumber $k\!\sim\!H^{-1}$ and a maximal growth rate $\Re[\Lambda]\!\sim\!f c_s/H$, in the presence of strong bimaterial contrast (e.g.~a deformable body sliding on top of a rigid one). This universal instability has been shown to be mediated by waveguide-like modes that are intrinsically coupled to the system height $H$. Furthermore, in has been shown that in the limit of large systems, $H\!\to\!\infty$, there exist --- in addition to quasi-static instability modes that are relevant at small sliding velocities~\cite{Rice2001} --- also fully dynamic instabilities. These are mediated by propagating plane-wave-like modes, featuring interesting directionality effects with respect to the sliding direction~\cite{Brener2016}. All in all, these examples show how the interplay between interfacial friction and bulk elasticity determines the stability of homogeneous frictional sliding.

\subsection{The onset of sliding motion: Creep patches and their stability}
\label{subsec:creep_patches}

\begin{figure}[ht]
  \centering
  \includegraphics[width=1\textwidth]{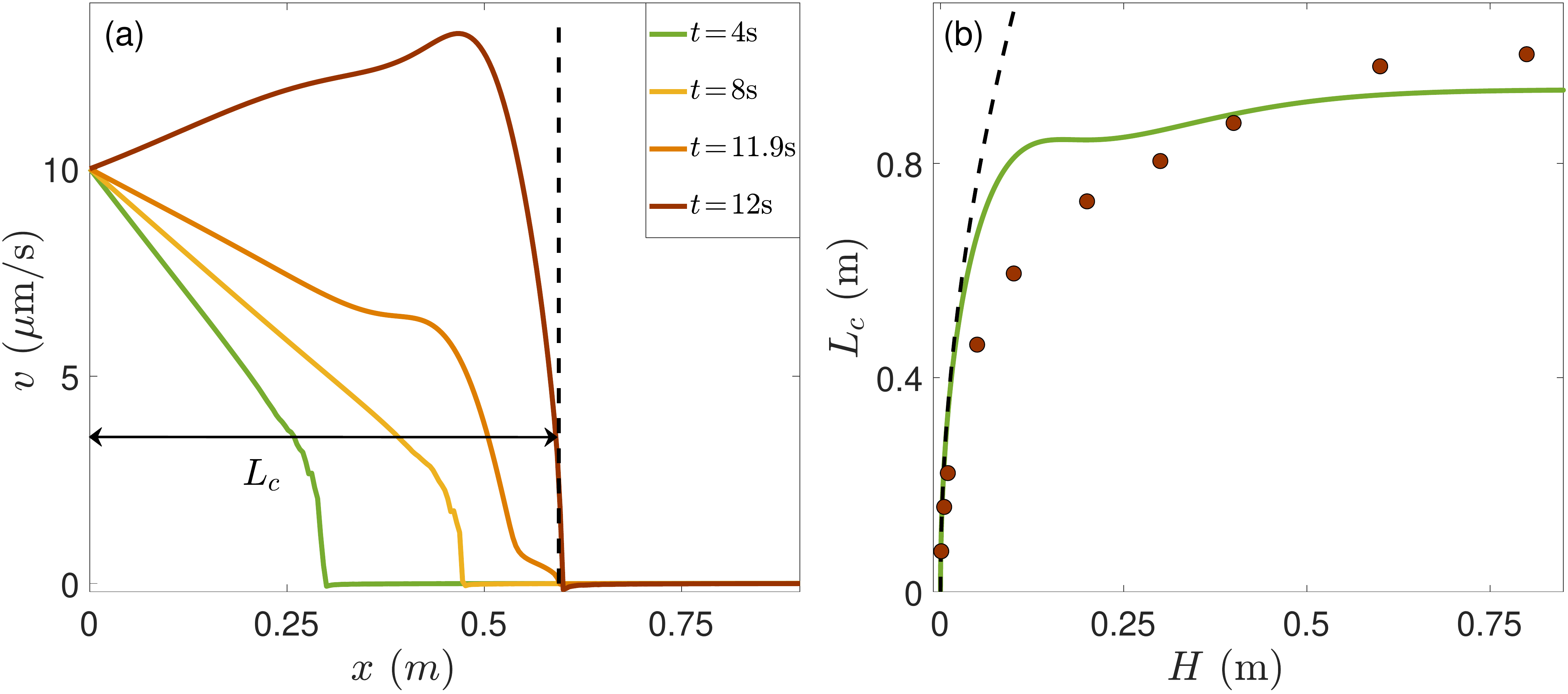}
  \caption{Loss of stability in transient, spatially inhomogeneous frictional dynamics under sideways loading (adapted from Fig.~3 of~\cite{Aldam2017a}). A long elastic body of height $H$, whose trailing edge ($x\!=\!0$) is pushed at a velocity of $10\mu$m/s, is considered. (a) A creep patch, manifested through different temporal snapshots of the slip velocity field $v(x,t)$ (see legend), is observed to propagate from the trailing edge into the interface as time progresses. When the sliding region reaches a critical size $L_c(H)$ (see text for details), the system loses stability and $v$ rapidly grows (note the time difference between the two upper curves). The subsequent dynamics feature a rupture front that rapidly propagates further into the interface ($x\!>\!L_c(H))$, which is not shown. (b) Dependence of critical length $L_c$ on system height $H$. The lines show analytical predictions for a 2D analysis (solid line) and a small-$H$ limit analysis (dashed line). The two agree for small $H$, while the former saturates at a finite value for large $H$. The circles show numerical results from and 2D finite-$H$ FEM simulations~\cite{Aldam2017a}. Reasonable and non-trivial agreement between the theory, which is based on homogeneous linear stability of steady sliding, and the numerics, based on transient inhomogeneous simulations, is demonstrated (see~\cite{Aldam2017a} for additional discussion).}
  \label{fig:patch}
\end{figure}

The discussion in the previous subsection focused on homogeneous sliding that results from homogeneous loading conditions. However, in many engineering applications and laboratory experiments the loading on a frictional system is inhomogeneous, e.g.~it is applied to the trailing edge of the bodies involved. To address this situation, we consider a body of height $H$ in frictional contact with an infinitely rigid substrate that is being pushed sideways, say at $x\=0$, with a velocity $v_{\rm d}$~\cite{Bar-Sinai2013,Aldam2017a}. The latter is assumed to be on the velocity-weakening branch of the steady-state friction curve in Fig.~\ref{fig:ss_friction}, i.e.~$\tfrac{df_{\rm ss}(v)}{d\log{v}}\!<\!0$ at $v_{\rm d}$. The inhomogeneity of this loading configuration, where at time $t\=0$ the slip velocity is $v\=v_{\rm d}$ at $x\=0$ and $v\=0$ for $x\!>\!0$, gives rise to the propagation of quasi-static creep patches, shown in Fig.~\ref{fig:patch}a based on a FEM calculation~\cite{Aldam2017a}. Creep patches progressively transfer stress and slip into the interface with increasing time $t$. Their quasi-static nature is explicitly revealed when the time dependence of their length, $l(t)$, is considered; in the small $H$ limit it takes the form $l(t)\!\sim\!\sqrt{H \,v_{\rm d} t}$ and in the limit of large $H$ it reads $l(t)\!\sim\! v_{\rm d} t$~\cite{Bar-Sinai2013}. In both limits, and also for intermediate $H$ values, $l(t)$ depends on $v_{\rm d}$ and in particular stops growing when the loading is halted, $v_{\rm d}\=0$.

Do the creep patches propagate indefinitely as long as $v_{\rm d}$ is maintained finite? The stability analysis of homogeneous sliding presented above suggests that this is not the case. While creep patches are characterized by inhomogeneous slip velocity distributions (cf.~Fig.~\ref{fig:patch}a), varying from $v(x\=0,t)\=v_{\rm d}$ to $v(x\!>\!l(t),t)\=0$ at any time $t$, the values themselves are predominantly on the velocity-weakening branch of the steady-state friction curve; overall, the physical situation is not expected to dramatically differ from that of a homogeneous patch characterized by an average slip velocity $\bar{v}(t)\=[l(t)]^{-1}\!\int_0^{l(t)}\!v(x,t)dx \propto v_{\rm d}$ that belongs to the velocity-weakening branch. Consequently, the stability analysis discussed above predicts that creep patches undergo a linear instability when their size surpasses the critical (minimal) length for instability, i.e.~when $l(t\=t_c)\=L_c(H)$. That is, when they become large enough to contain the smallest unstable mode. This prediction is tested in Fig.~\ref{fig:patch}a and is fully supported by the FEM calculation, i.e.~when $l(t)$ surpasses $L_c(H)$ (for the $H$ employed in this calculation), an instability characterized by an exponential growth of slip velocities is observed.

To test this prediction over a broad range of $H$ values, $L_c(H)$ has been theoretically calculated using the homogeneous linear stability analysis discussed in Sect.~\ref{subsec:LSA}~\cite{Aldam2017a}. Note, though, that the present problem is more complicated than the case discussed in Sect.~\ref{subsec:LSA} due to the existence of bimaterial contrast, i.e.~the present problem involves a deformable body sliding on top of a rigid one. The scaling properties of the solutions reported on in Sect.~\ref{subsec:LSA} are nevertheless preserved, i.e.~$L_c(H)\!\sim\!\sqrt{H}$ for small $H$ and $L_c(H)\!\to\!\hbox{const.}$ for large $H$, as shown in Fig.~\ref{fig:patch}b. The comparison between the size in which creep pathes lose their stability, measured in dynamical simulations of propagating creep patches for various $H$ values, and the theoretical prediction for $L_c(H)$ is also shown in Fig.~\ref{fig:patch}b. The dynamics of creep patches and their accompanying instabilities determine the onset of global frictional motion, occurring when slip first reaches the leading edge of the body. These rich and realistic spatiotemporal dynamics crucially depend on the interplay between interfacial friction and bulk elasticity (as well as other related analyses in the literature, e.g.~\cite{Lapusta2003,Rubin2005,Ampuero2008,Kammer2012,Viesca2016b}), and are qualitatively different from the description of the onset of frictional motion using spring-block models that lack spatially-extended degrees of freedom. In fact, the transfer of slip to the leading edge of the body also involves dynamically propagating modes that are excited by the instabilities of creep patches (e.g.~the one shown in Fig.~\ref{fig:patch}a), not discussed up until now. Such propagating frictional modes are discussed next.

\subsection{Propagating frictional modes}
\label{subsec:propagating_modes}

Propagating frictional modes play crucial roles in the dynamics of frictional systems~\cite{Svetlizky2019,Ben-Zion2001,Scholz2002,Rubinstein2004,Rubinstein2007,Kammer2012,Svetlizky2014}. For example, following the previous subsection, propagating frictional modes that are generated by instabilities of creep patches play important roles in the onset of global frictional motion, relevant to a wide variety of engineering systems. Our goal in this subsection is to understand how such propagating frictional modes emerge from the interplay between interfacial friction and bulk elasticity and what forms can they take. To this aim, we consider a frictional system subjected to homogeneous shear stress and normal (compressive) stress loading, $\tau_0$ and $\sigma_0$ respectively (cf.~Fig.~\ref{fig:sketch}).

Suppose then that $\tau_0$ is larger than $\tau_{\rm min}\!\equiv\!\sigma_0 f_{\rm min}$, where $f_{\rm min}$ corresponds to the minimum of the steady-state friction curve in Fig.~\ref{fig:ss_friction}. Under these conditions, there exist $3$ spatially homogeneous fixed-points (already mentioned above) that correspond to the intersection points of the $\tau_0/\sigma_0$ line with $f_{ss}(v)$, cf.~Fig.~\ref{fig:ss_friction}. The first and last intersection points, on the velocity-strengthening branches, are stable fixed-points from the perspective of the friction law, while the intermediate one, on the velocity-weakening branch, is an unstable fixed-point (it has been discussed in the context of creep patches in Sect.~\ref{subsec:creep_patches}). Generally speaking, frictional modes that propagate at a steady-state speed $c$ spatially connect the two stable fixed-points. The steady nature of these propagating objects imply that they can be described in a co-moving frame defined through $\xi\!\equiv\!x-ct$, and consequently that $\partial_t\!=\!-c\partial_\xi$ and $\partial_x\!=\!\partial_\xi$ such that all of the relevant equations can be expressed in terms of the co-moving coordinate $\xi$.

We discuss below $3$ types of steady-state propagating frictional modes. First, we consider modes that connect the very low-$v$ stable fixed-point ($v\!\approx\!0$, termed the `stick state') with the high-$v$ stable fixed-point ($v\!>\!0$, the `sliding/slipping state'). If the latter is realized at $\xi\!\to\!-\infty$ and the former at $\xi\!\to\!\infty$, the modes propagate from left to right with $c\!>\!0$, and are characterized by a lengthscale $\ell$ around $\xi\=0$ over which the `stick state' changes to the `sliding/slipping state'. These propagating frictional modes are sometimes termed frictional rupture fronts. Second, when the stick state invades the sliding state, we obtain healing (`inverse') frictional fronts. Finally, when the stick state exists both at $\xi\!\to\!\pm\infty$, with a spatially-compact region of $v\!>\!0$ around $\xi\=0$, we obtain slip pulses. As will be explained below, slip pulses can be constructed from the interaction of rupture and healing frictional fronts (we note in passing that there exists a related frictional mode, composed of a periodic train of pulses~\cite{Brener2005,Shi2008,Heimisson2019}, which is not discussed here). In all cases, propagating frictional modes involve the self-selection of both the characteristic transition length $\ell$ and the propagation speed $c$. Our goal below is to understand how they are determined by the interplay between interfacial friction and bulk elasticity.

As a first step, we show that $\ell$ and $c$ satisfy a generic scaling relation. To see this, note that the time it takes for a propagating frictional mode to travel over a given point on the interface is $\ell/c$. During this time, the slip accumulated at this point scales as $v(\ell)\,\ell/c\!\sim\!v_{_{\rm p}}\,\ell/c$, where we estimated the slip velocity in the transition region $v(\ell)$ by the peak (maximal) slip velocity $v_{_{\rm p}}$. Finally, since the accumulated slip scales with the typical slip distance $D$ (while the accumulated slip itself is quite significantly larger than $D$~\cite{Cocco2002}), we obtain
\begin{equation}
\frac{c}{\ell} \sim \frac{v_{_{\rm p}}}{D} \ .
\label{eq:ell_c_scaling}
\end{equation}
This relation can be somewhat more formally rationalized using Eq.~\eqref{eq:phidot} (the same reasoning applies to Eq.~\eqref{eq:tauel}). Note that the relation between $\ell$ and $c$ in Eq.~\eqref{eq:ell_c_scaling} is independent of the loading conditions and dimensionality as a scaling relation, though an implicit dependence emerges through $v_{_{\rm p}}\=v_{_{\rm p}}(\tau_0/\sigma_0,H)$, as will be discussed below.
\begin{figure}[ht]
  \centering
    \includegraphics[width=\textwidth]{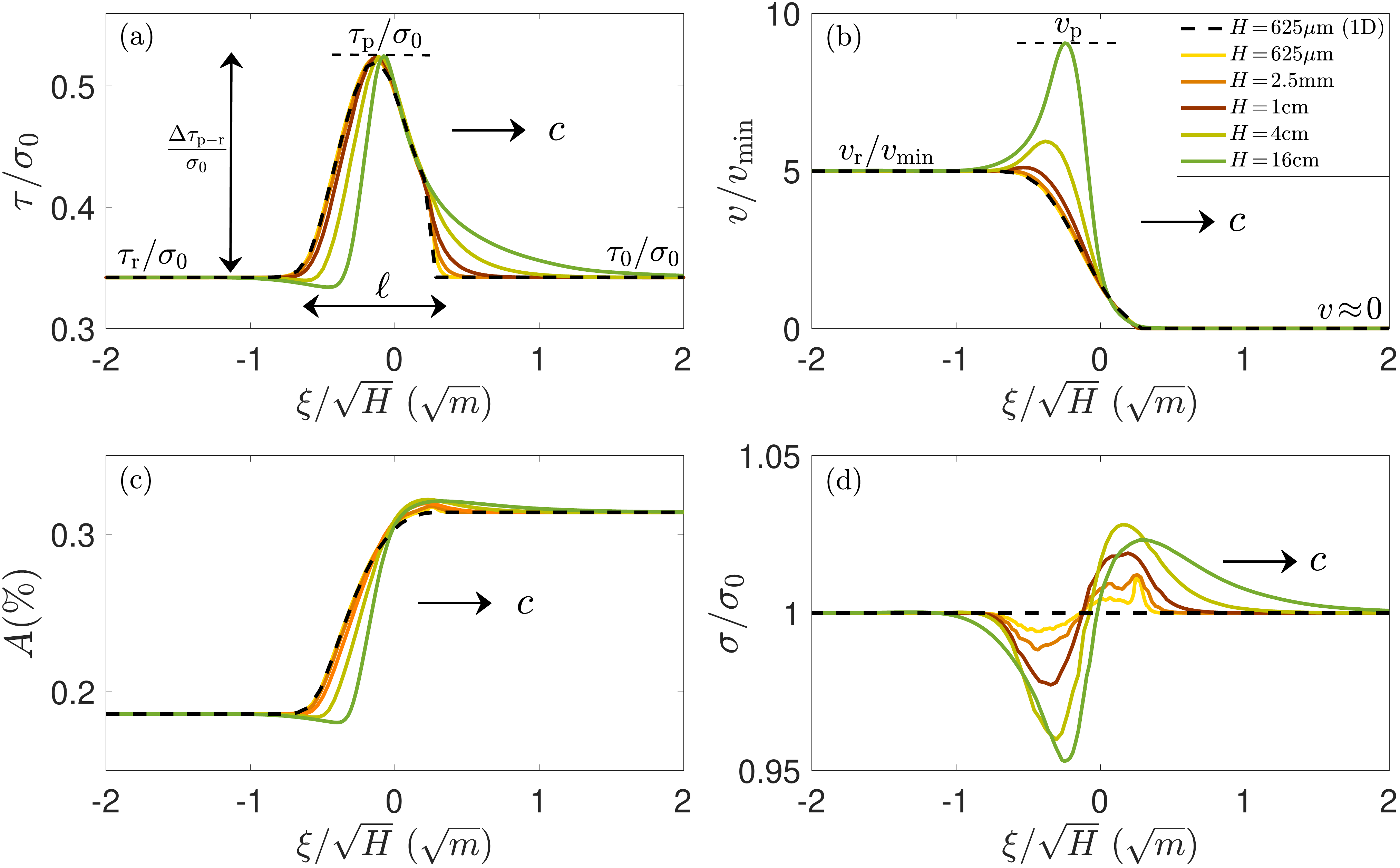}
  \caption{Spatial profiles of steady-state rupture fronts for varying $H$ and fixed $\tau_0/\sigma_0$. The solid lines correspond to 2D treadmill FEM calculations (see~\hyperref[sec:treadmill]{Appendix C} for details and the legend in panel (b) for the $H$ values used). The dashed line corresponds to a quasi-1D small-$H$ limit calculation (for the smallest $H$). The agreement between the two different calculations in the small-$H$ limit is evident. Shown are the (a) shear stress, (b) slip velocity, (c) real contact area and (d) normal stress (all properly normalized). All fields are shown as a function of the co-moving coordinate $\xi$, rescaled by $\sqrt{H}$ in order to highlight the collapse of all curves for small $H$, as predicted by Eq.~\eqref{eq:c_ell}. At larger $H$, fronts develop a peak in the slip velocity, cf.~panel (b), and a variation in the normal stress, cf.~panel (d). All quantities defined in the text are marked on the figure, note in particular the transition length $\ell$ and the propagation speed $c$ (these, like all other quantities marked, depend on $H$).}
  \label{fig:2D}
\end{figure}

To see how these physical quantities emerge in realistic frictional systems, we first discuss steady-state propagating rupture modes. To that aim, we consider a long deformable body of height $H$ in frictional contact with a rigid substrate, subjected at $y\!=\!H$ to homogeneous shear and normal stress, $\tau_0$ and $\sigma_0$ respectively (cf.~Fig.~\ref{fig:sketch}). We then obtain steady-state propagating rupture modes using a treadmill FEM routine detailed in~\hyperref[sec:treadmill]{Appendix C}. The results, demonstrating steady-state rupture modes propagating from left to right for a broad range of $H$ values, are presented in Fig.~\ref{fig:2D}. In panel (a) we present $\tau(\xi)/\sigma_0$, focusing on the transition region of size $\ell$ that clearly emerges near $\xi\=0$. It is observed that far ahead of the transition region $\tau(\xi)$ simply equals to the applied shear stress $\tau_0$, as expected. Far behind the transition region, $\tau(\xi)$ approaches a residual stress $\tau_{\rm r}$, which identifies here with the loading shear stress $\tau_{\rm r}\!\to\!\tau_0$; this is expected since the two stable fixed-points that the rupture mode is connecting feature the same stress $\tau_0$ (cf.~the intersection points marked by black squares in Fig.~\ref{fig:ss_friction}). The fact that the residual and applied shear stresses are identical for steady-state rupture modes will be further discussed below. Finally, in the transition region, $\tau(\xi)$ goes through a maximum, which we denote by $\tau_{_{\rm p}}$.

The corresponding slip velocity $v(\xi)/v_{\rm min}$ is shown in Fig.~\ref{fig:2D}b. The residual slip velocity left behind the propagating rupture mode, $v_{\rm r}$, corresponds to the high-$v$ stable fixed-point in Fig.~\ref{fig:ss_friction}, $\tau_0/\sigma_0\=f_{\rm ss}(v_{\rm r})$, while far ahead of it the slip velocity is vanishingly small, $v\!\approx\!0$ (formally also a solution of $\tau_0/\sigma_0\=f_{\rm ss}(v)$). In between, the two are smoothly connected over a lenthscale $\ell$, where the connection is monotonic for small $H$ and nonmonotonic for larger $H$, going through a maximum denoted by $v_{_{\rm p}}$. When the transition is monotonic, we simply have for the maximal slip velocity $v_{_{\rm p}}\=v_{_{\rm r}}$. In panel (c) the normalized real contact area $A(\xi)$ is presented, directly justifying the term `rupture front' as a low $A$ value state (`partially ruptured contact') is observed to invade a high $A$ value state (reference real contact area). Finally, in panel (d) we present $\sigma(\xi)/\sigma_0$, which is simply unity for small $H$ (since $\delta\!{\cal F}_{_{\!\sigma}}\=0$ in this limit) and deviates from unity as $H$ increases due to the bimaterial effect manifested in $\delta\!{\cal F}_{_{\!\sigma}}\!\ne\!0$ (recall that we are considering a deformable body sliding on top of a rigid one).

To gain deeper insight into $\ell$ and $c$, going beyond the scaling relation in Eq.~\eqref{eq:ell_c_scaling}, we follow the scaling analysis of~\cite{Bar-Sinai2012} in the small-$H$ limit ($H\!\ll\!\ell$), to obtain
\begin{eqnarray}
\label{eq:c_ell}
\ell\!\sim\!\sqrt{\frac{\bar{\mu}\,H\,D}{\Delta\tau_{_{\rm p-r}}}}\qquad~\hbox{and}\qquad~
c \sim v_{_{\rm p}}(\tau_0/\sigma_0,H)\sqrt{\frac{\bar{\mu}\,H}{\Delta\tau_{_{\rm p-r}}\,D}}\sim v_{_{\rm r}}(\tau_0/\sigma_0)\sqrt{\frac{\bar{\mu}\,H}{\Delta\tau_{_{\rm p-r}}\,D}} \qquad\hbox{for}\qquad H\!\ll\!\ell \ ,
\end{eqnarray}
where the so-called dynamic stress drop is defined as $\Delta\tau_{_{\rm p-r}}\!\equiv\!\tau_{_{\rm p}}-\tau_{_{\rm r}}$. Note that strongly inertial effects, which emerge when $c$ approaches the elastic wave speeds, have been neglected, that we used the fact that $v_{_{\rm p}}(\tau_0/\sigma_0,H)\=v_{_{\rm r}}(\tau_0/\sigma_0)$ in the small-$H$ limit, and that $\ell$ and $c$ indeed satisfy the scaling relation in Eq.~\eqref{eq:ell_c_scaling}. The relations in Eq.~\eqref{eq:c_ell} clearly demonstrate how $\ell$ and $c$ emerge from the interplay between interfacial friction and bulk elasticity. The prediction $\ell\!\sim\!\sqrt{H}$ is tested in Fig.~\ref{fig:2D} through rescaling $\xi$ by $\sqrt{H}$. It is observed that for small $H$, but over a large range (here $H$ varies $16$-fold, from $625 \mu$m to $1$cm), the fields nearly collapse on $H$-independent curves, verifying the $\ell\!\sim\!\sqrt{H}$ prediction. For yet larger $H$ values, some deviations are observed, as will be discussed next. The scaling prediction for $c$ in Eq.~\eqref{eq:ell_c_scaling} will also be tested below.

With increasing $H$, we expect the small-$H$ predictions in Eq.~\eqref{eq:c_ell} to break down gradually, eventually being replaced by their large-$H$ counterparts ($H\!\gg\!\ell$). The procedure to obtain the latter from Eq.~\eqref{eq:c_ell} is in principle straightforward; one should just replace $H$ in the first relation in Eq.~\eqref{eq:c_ell} by $\ell$ and solve for the latter to obtain $\ell\!\sim\!\mu\,D/\Delta\tau_{_{\rm p-r}}$ (the small-$H$ shear modulus $\bar{\mu}$ is replaced here by $\mu$). Then the latter is used to replace $H$ in the second relation in Eq.~\eqref{eq:c_ell}, leading to $c\!\sim\!v_{_{\rm p}}\mu/\Delta\tau_{_{\rm p-r}}$. The only remaining issue is the $H$-dependence of $v_{_{\rm p}}(\tau_0/\sigma_0,H)$, which is clearly observed for the largest $H$ values used in Fig.~\ref{fig:2D}b. This pronounced increase of $v_{_{\rm p}}$ with $H$ is nothing but a signature of the build-up of the famous square root singularity of crack-like objects~\cite{Freund1998}, which is absent in the quasi-one-dimensional limit of small $H$. Consequently, using conventional fracture mechanics in long systems (recall that we assumed $L\!\to\!\infty$, i.e.~our system features a `strip geometry' $H\!\ll\!L$)~\cite{Freund1998}, the residual slip velocity $v_{_{\rm r}}$ is amplified in the transition region according to $v_{_{\rm p}}(\tau_0/\sigma_0,H)\!\sim\! v_{_{\rm r}}(\tau_0/\sigma_0)\sqrt{H/\ell}$. Taken together, we obtain for the large-$H$ limit the following predictions
\begin{eqnarray}
\label{eq:c_ell2D}
\ell\!\sim\!\frac{\mu\,D}{\Delta\tau_{_{\rm p-r}}} \qquad\qquad\hbox{and}\qquad\qquad
c \sim v_{_{\rm r}}(\tau_0/\sigma_0)\sqrt{\frac{\mu\,H}{\Delta\tau_{_{\rm p-r}}\,D}} \qquad\qquad\hbox{for}\qquad\qquad \ell\!\ll\!H \ .
\end{eqnarray}

\begin{figure}[ht]
  \centering
    \includegraphics[width=0.9\textwidth]{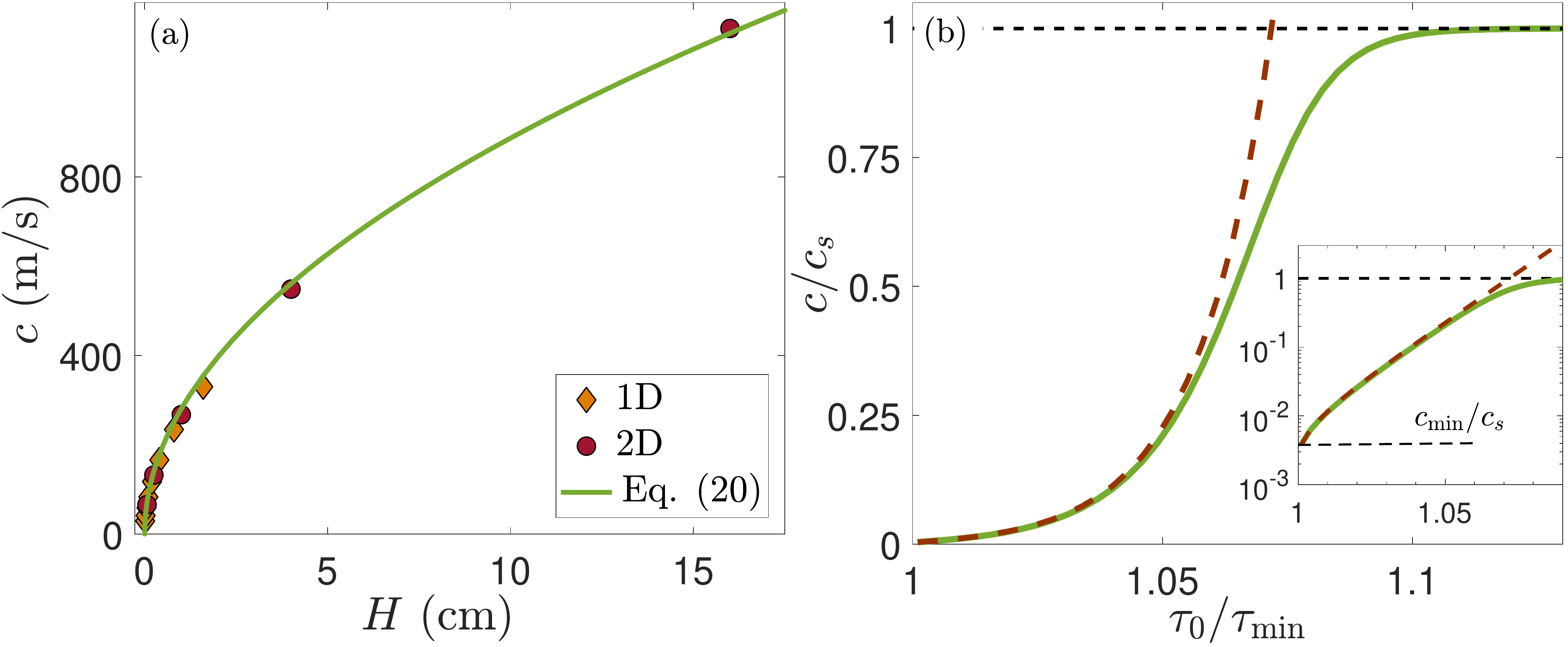}
  \caption{Steady-state rupture front speeds. (a) The propagation speed $c$, corresponding to the 2D steady-state rupture fronts shown in Fig.~\ref{fig:2D}, as a function of $H$ (circles). Results using quasi-1D calculations for various $H$ values are added (diamonds). The solid line is the theoretical prediction for $c(H)$ in Eqs.~\eqref{eq:c_ell}-\eqref{eq:c_ell2D}, with a prefactor of order unity, demonstrating favorable agreement with the numerical results. Note that both $c$ and $H$ are dimensional, and that the presented results are obtained in the quasi-static regime where Eqs.~\eqref{eq:c_ell}-\eqref{eq:c_ell2D} are valid (the wave speed in the 2D calculations has been assigned a large value, see~\hyperref[sec:treadmill]{Appendix C}). (b) Rupture propagation speed as a function of loading stress $\tau_0$ for fixed $H$ in fully inertial quasi-1D calculations (solid line), highlighting inertial effects. The quasi-static approximation (dashed line) agrees with the inertial solutions as long as $c$ is sufficiently smaller than the wave speed $c_s$, while strong deviations are observed when $c$ saturates at $c_s$. Inset: same data in logarithmic vertical axis (axis labels are not shown again), highlighting the existence of a finite minimal propagation speed $c_{\rm min}$ and the exponential growth of $c$ with $\tau_0$ in the quasi-static regime.}
  \label{fig:spectrum}
\end{figure}

This analysis seems to suggest that the propagation speed $c$ follows the scaling relation $c\!\sim\!\sqrt{H}$ in both the small and large $H$ limits, cf.~Eqs.~\eqref{eq:c_ell}-\eqref{eq:c_ell2D}. While at present we cannot comprehensively test this prediction, simply because very large $H$ values are computationally inaccessible, we can consider the available results of Fig.~\ref{fig:2D}, where deviations from the small-$H$ limit are observed (e.g.~the lack of collapse of the curves shown in panel (a) for the largest $H$ values used and the appearance of a significant overshoot in the slip velocity profiles shown in panel (b) there). Consequently, we plot in Fig.~\ref{fig:spectrum}a $c(H)$ for the results presented in Fig.~\ref{fig:2D}. It is observed that the scaling prediction $c\!\sim\!\sqrt{H}$ is accurately followed up to the largest $H$ value considered here. Upon further increasing $H$, $c$ cannot increase without bound and we expect inertial effects to intervene, leading to the saturation of $c$ at the elastic wave speed. As just explained, probing this limit using a treadmill FEM routine is practically impossible due to the computational costs associated with large $H$ values. Yet, we do demonstrate the inertial saturation effect by considering the dependence of $c$ on the applied shear stress $\tau_0$, a dependence that has not been addressed up to now, using numerical calculations in the small-$H$ limit. In fact, note that results of such calculations (using ${\cal F}_{_{\!\tau}}$ of Eq.~\eqref{eq:1d}) and efficient numerics~\cite{Bar-Sinai2012,Bar-Sinai2013,Bar-Sinai2015a}) have already been included in Fig.~\ref{fig:2D} and Fig.~\ref{fig:spectrum}a, demonstrating perfect agreement with the small-$H$ FEM results. According to Eqs.~\eqref{eq:c_ell}-\eqref{eq:c_ell2D}, the $\tau_0$ dependence of $c$ is fully contained in $v_{_{\rm r}}(\tau_0/\sigma_0)$, which is a property of the high-$v$ velocity-strengthening branch of the steady-state friction curve (determined through $\tau_0/\sigma_0\=f_{\rm ss}(v_{\rm r})$).

The `spectrum' of propagation speeds $c(\tau_0)$ is shown in Fig.~\ref{fig:spectrum}b. Note that $c$ is well-defined for $\tau_0\!>\!\tau_{\rm min}$, as only in this regime steady-state rupture fronts exist~\cite{Bar-Sinai2012}. Interestingly, as discussed extensively in~\cite{Bar-Sinai2012}, in the limit $\tau_0\!\to\!\tau_{\rm min}$ the propagation speed $c$ attains a {\em finite} value that is much smaller than the elastic wave speed $c_s$ (cf.~inset), a result that might be related to the long-debated problem of slow earthquakes/slip phenomena~\cite{Rubinstein2004,Ben-David2010a,Kaproth2013,Burgmann2018}. As $\tau_0$ is increased away from $\tau_{\rm min}$, the propagation speed $c$ increases according to the theoretical prediction of Eq.~\eqref{eq:c_ell}, which corresponds to the dashed line. However, inasmuch as $c$ cannot increase indefinitely with $H$, it cannot also increase indefinitely with $\tau_0$. In particular, we expect that once $c$ becomes comparable to $c_s$, inertial effects become dominant and $c\!\to\!c_s$. This expectation is fully supported by the results of Fig.~\ref{fig:spectrum}b. In the inertia-dominated regime other effects that are not discussed here emerge. For example, the characteristic length $\ell$ undergoes relativistic (Lorentz) contraction that follows a $\sqrt{1-c^2/c_s^2}$ (in the large $H$ limit, $c_s$ is replaced by the Rayleigh wave speed~\cite{Freund1998,Svetlizky2014}).

\begin{figure}[ht]
  \centering
    \includegraphics[width=\textwidth]{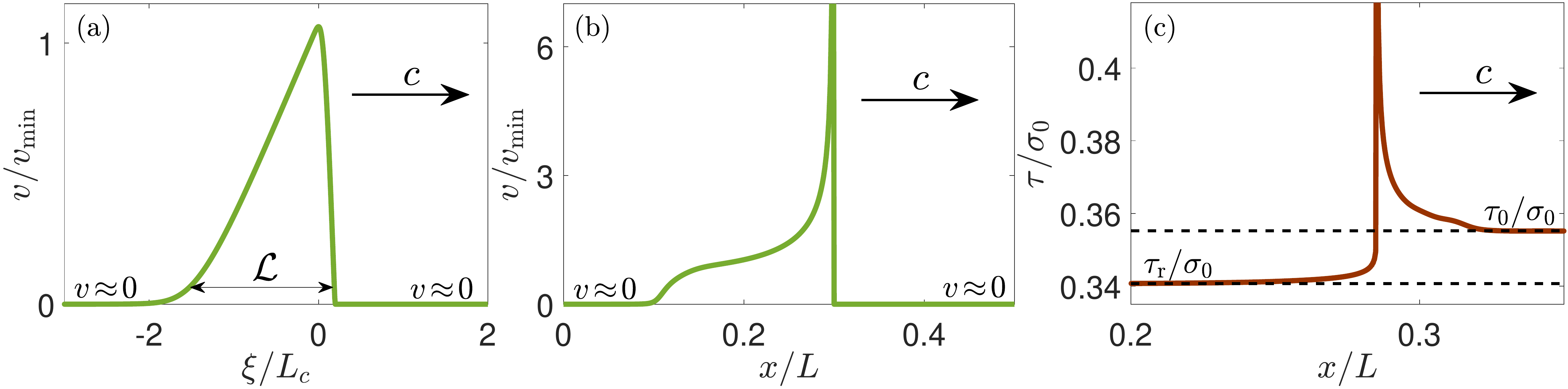}
  \caption{(a) A steady-state slip pulse of characteristic size ${\cal L}$ obtained in the small-$H$ limit using ${\cal F}_{_{\!\tau}}$ of Eq.~\eqref{eq:1d} (adapted from Fig.~2 in~\cite{Brener2018}). Shown is the slip velocity profile as a function of the co-moving coordinate $\xi$ used in steady-state calculations (normalized by $L_c$ of Eq.~\eqref{eq:Lc}). (b) A sustained slip pulse obtained through a spectral boundary integral method calculation in the 2D infinite-$H$ limit using ${\cal F}_{_{\!\tau}}$ of Eq.~\eqref{eq:infiniteH_dynamic} (extracted from supplementary movie~S4 in~\cite{Brener2018}). The curve is truncated in the vertical axis for visual clarify. (c) A non-steady 2D infinite-$H$ rapid rupture front generated through a spectral boundary integral method calculation (adapted from Fig.~3 in~\cite{PartI}). The curve is truncated in the vertical axis for visual clarify. For all panels, see text for additional details.}
  \label{fig:front_pulse}
\end{figure}

As mentioned above, one can consider in addition to frictional rupture fronts also frictional healing fronts. In principle, the discussion and procedures presented above in relation to rupture fronts can be applied to healing fronts as well, where the stick state invades the sliding state instead of vice versa. Here we focus on just one aspect of the results concerning healing fronts, the one that addresses the spectrum of propagation speeds $c(\tau_0)$. The major difference compared to the results presented in Fig.~\ref{fig:spectrum}b, where $c(\tau_0)$ for rupture fronts is shown to be an increasing function of $\tau_0$, is that for healing fronts $c(\tau_0)$ is a decreasing function of $\tau_0$~\cite{Putelat2017,Brener2018}. This implies that the two propagation speed spectra intersect at some loading level, say $\tau^*$. The latter is of physical significance because it is related to the third type of steady-state propagating frictional modes mentioned above, i.e.~to slip pulses (sometimes also termed self-healing pulses)~\cite{Heaton1990,Noda2009,Putelat2017}. The point is that when the system is driven at $\tau_0\=\tau^*$, rupture and healing fronts propagate at the same speed and hence can be superimposed without interaction to form a steady-state pulse of infinite width. As $\tau_0$ is increased above $\tau^*$, the two fronts interact, leading to pulses of finite width ${\cal L}(\tau_0)$, with ${\cal L}\!\to\!\infty$ for $\tau_0\!\to\!\tau^*$. An example of such a finite-width propagating slip pulse in the small-$H$ limit (i.e.~using ${\cal F}_{_{\!\tau}}$ of Eq.~\eqref{eq:1d}) is presented in Fig.~\ref{fig:front_pulse}a~\cite{Brener2018}.

Up until now, we focused on steady-state propagating frictional modes, i.e.~on solutions that depend only on the co-moving coordinate $\xi\=x-ct$.
These modes can be also studied in fully dynamic numerical simulations in which steady-state conditions are not enforced. A snapshot from such a simulation in the infinite-$H$ limit (i.e.~using ${\cal F}_{_{\!\tau}}$ of Eq.~\eqref{eq:infiniteH_dynamic}) is shown in Fig.~\ref{fig:front_pulse}b~\cite{Brener2018}. The figure presents a slip pulse that appears to propagate over long distances without appreciably changing its shape and propagation speed. Such pulses are sometimes termed sustained pulses~\cite{Brener2018}, though determining whether they are truly steady-state objects or not is difficult based on dynamic simulations. The quite pronounced differences in the shape of the slip pulses presented in Figs.~\ref{fig:front_pulse}a-b are related to dimensionality (i.e.~the former is obtained in the quasi-1D limit of small $H$ and the latter in the 2D infinite-$H$ limit).

Looking for steady-state solutions by using the co-moving coordinate $\xi\=x-ct$ has other implications. Most notably, by so doing one excludes from the discussion the possibility that some of the obtained steady-state solutions are in fact unstable. It turns out that small-$H$ steady-state pulse solutions, an example of which is presented in Fig.~\ref{fig:front_pulse}a, are actually unstable~\cite{Brener2018}. In particular, it has been shown that small-$H$ steady-state pulse solutions can be used to sort out perturbations based on their spatial extent; suppose that a homogeneous state corresponding to the stick fixed-point of the steady-state friction curve at a given $\tau_0\!>\!\tau^*$ is introduced with a slip velocity perturbation. If this perturbation is narrower than the characteristic width of the corresponding steady-state slip pulse (i.e.~its spatial extent is smaller than ${\cal L}(\tau_0)$), then it decays in the subsequent dynamics (where steady-state is not enforced). On the other hand, perturbations characterized by a spatial extent larger than ${\cal L}(\tau_0)$ are amplified and develop into propagating rupture fronts~\cite{Brener2018}.

The unstable nature of small-$H$ steady-state slip pulses has led to the idea that they might in fact serve as critical nuclei for rapid slip propagation mediated by rupture fronts, in analogy to first order phase transitions. That is, it has been suggested that slip pulses, and in particular their characteristic size ${\cal L}(\tau_0)$, determine the minimal size of perturbations needed for a homogeneous stick state to develop rupture fronts that bring the system to the high-$v$ velocity-strengthening fixed-point for the same $\tau_0$. This idea has also been tested in the 2D infinite-$H$ limit, where the critical nucleus size ${\cal L}(\tau_0)$ has been identified (obviously this ${\cal L}(\tau_0)$ is larger than the width of the sustained pulses observed at the same $\tau_0$, cf.~Fig.~\ref{fig:front_pulse}b). An example of a rapid rupture front emerging from introducing a perturbation larger than ${\cal L}(\tau_0)$ into a homogeneous stick state, in the infinite-$H$ limit, is shown in Fig.~\ref{fig:front_pulse}c. This phase transition-like nucleation process of rupture fronts at locked-in (i.e.~stick state) frictional interfaces is qualitatively different from the linear frictional instability nucleation process at velocity-weakening interfaces, discussed in Sect.~\ref{subsec:creep_patches} (also compare to the nucleation scenario proposed in~\cite{Uenishi2003}).

The 2D infinite-$H$ rapid rupture front shown in Fig.~\ref{fig:front_pulse}c highlights another interesting bulk elasticity effect in frictional systems, which concludes our discussion. Comparing the rupture front in Fig.~\ref{fig:2D}a to the one in Fig.~\ref{fig:front_pulse}c reveals a qualitative difference; the former leaves behind it a residual stress $\tau_{\rm r}$ that equals the applied shear stress $\tau_0$ (achieved ahead of the front), while in the latter we observe $\tau_{\rm r}\!<\!\tau_0$. Put differently, the former features a vanishing stress drop $\Delta\tau_{_{\rm 0-r}}\!\equiv\!\tau_0-\tau_{_{\rm r}}\!\to\!0$ ($\Delta\tau_{_{\rm 0-r}}$ should be distinguished from the dynamic stress drop $\Delta\tau_{_{\rm p-r}}$ discussed above), while the latter features a finite stress drop $\Delta\tau_{_{\rm 0-r}}$. The existence of stress drops has important implications for frictional rupture~\cite{PartI,PartII,Svetlizky2019,Lu2010}. The origin of this qualitative difference is that ${\cal F}_{_{\!\tau}}$ of Eq.~\eqref{eq:infiniteH_dynamic}, whose value behind the rupture fronts is simply $\Delta\tau_{_{\rm 0-r}}$ (cf.~Eq.~\eqref{eq:coupling}), is finite for transient, out of steady-state rapid rupture (due to both the radiation-damping term and the long-ranged interaction term $s(x,t)$ in Eq.~\eqref{eq:infiniteH_dynamic}), while it vanishes under steady-state conditions in finite-$H$ systems~\cite{PartI}. These findings demonstrate that even a quantity such as the residual stress $\tau_{_{\rm r}}$, which is traditionally regarded as a property of interfacial friction law~\cite{Ida1972,Palmer1973}, is in fact strongly affected by bulk elasticity and bulk dynamics.

\section{Concluding remarks}
\label{sec:summary}

Frictional systems pose great conceptual, physical and mathematical challenges, with far-reaching scientific and technological implications. A frictional system is generically composed of a low-dimensional part --- the contact interface --- and a higher-dimensional part --- the bulks that surround the interface. The major theme of this feature paper is that in order to understand the behavior of frictional systems one has to take into account both the interfacial physics that characterizes the frictional contact and the bulk-mediated interactions between different parts of the interface, and that this inherent interface-bulk coupling is indispensable.

In the context of macroscopic interfacial constitutive laws, we discussed the widely-used rate-and-state friction modeling framework. We highlighted the power of this coarse-grained approach and also some of its limitations, and then proposed physically-motivated modifications to overcome these limitations. This part of the paper emphasized the generic properties of interfacial constitutive relations, where the discussion focused on spatially-extended, multi-contact dry frictional interfaces. We did not explicitly take into account third-body effects, wear or various lubrication processes that may give rise to additional important and interesting interfacial physics. It is worthwhile mentioning that the steady-state friction curve of Fig.~\ref{fig:ss_friction} features all of the salient properties of the Stribeck curve~\cite{Woodhouse2015}, which generically describes fluid-lubricated contact interfaces. Consequently, we expect some aspects of our results to be relevant also to lubricated interfaces, offering some unifying concepts.

In the context of bulk-mediated interactions, we focused on a linear elastodynamic bulk constitutive relation, highlighting how macroscopic geometry and material contrast affect the range and nature of the emerging spatiotemporal interactions. While we believe that linear elastodynamics is generally relevant and applicable to frictional systems, other bulk physics might be relevant and important in various situations. The latter may include bulk viscoelasticity (e.g.~\cite{Radiguet2013,Radiguet2015,Allison2017}), poroelasticity (i.e.~the interaction between fluid flow and solid deformation in porous materials, e.g.~\cite{Dunham2008,Heimisson2019}), plasticity and distributed/off-fault damage (e.g.~\cite{Shi2010}), which can affect the overall behavior of some frictional systems.

Finally, we have shown how basic phenomena in frictional systems --- including instabilities of homogeneous sliding, the onset of sliding motion and a wide variety of propagating frictional modes (e.g.~rupture fronts, healing fronts and slip pulses) --- emerge from the inherent interplay of interfacial and bulk physics. We believe that these phenomena are generic and essential for understanding the behavior of a broad range of frictional systems. We therefore hope that the approach discussed and advocated in this feature paper will be useful for solving problems in fields ranging from engineering tribology to geophysics.\\

{\bf Acknowledgements} E. B. acknowledges support from the Israel Science Foundation
(Grant No.~295/16), from the William Z. and Eda Bess Novick Young Scientist Fund and from the Harold Perlman Family. Y.B.S.~acknowledges support from the James S.~McDonnell post-doctoral fellowship for the study of complex systems.

\setcounter{secnumdepth}{0}
\section{Appendices}

The goal of the appendices is to provide additional technical details about the results presented above, with a special focus on results that have not appeared elsewhere. The latter include the small stress frictional dynamics summarized in Fig~\ref{fig:IE}b, and the steady-state frictional modes results presented in Fig.~\ref{fig:2D} and Fig.~\ref{fig:spectrum}a, involving treadmill FEM calculations.

\setcounter{equation}{0}
\renewcommand{\theequation}{A\arabic{equation}}
\subsection{Appendix A: Summary of the mathematical formulation of the extended friction law}
\label{sec:math_friction}

The extended RSF friction law discussed above is formulated in Eqs.~\eqref{eq:KV},~\eqref{eq:tauel} and~\eqref{eq:A}, where the real contact area $A(\phi)$ of Eq.~\eqref{eq:A} is used in Eqs.~\eqref{eq:tauel} and~\eqref{eq:RSFsinh}, and the latter is interpreted as $\tau^{\rm vis}(\phi,v)$ in Eq.~\eqref{eq:KV}. Finally, the function $g(\cdot)$ is introduced in Eq.~\eqref{eq:phidot} as done in Eq.~\eqref{eq:tauel}. Collecting all of these equations, we obtain
\begin{equation}
f(\phi,v,\tau^{\rm el})\!=\!\frac{\tau^{\rm el} + \tau^{\rm vis}(\phi,v)}{\sigma}\!=\!\left[\frac{\tau^{\rm el}}{\sigma} +\frac{k_B T}{\Omega\,\sigma_{\hbox{\tiny H}}}\left[1\!+\!b\log\left(1+\frac{\phi}{\phi^*}\right)\right]\sinh^{-1}\!\!\left(\frac{v}{2\,v_c}\,\exp\left[\frac{E_0}{k_B T}\right]\right)\right]
\label{eqA:RSFsinh} \ ,
\end{equation}
\begin{equation}
\dot \phi =1- \frac{\phi|v|}{D}g(\cdot) \qquad\qquad\hbox{and}\qquad\qquad  \dot\tau^{\rm el}= \frac{\mu_0\,\sigma}{h\,\sigma_{\hbox{\tiny H}}}\left[1+b\log\left(1+\frac{\phi}{\phi^*}\right)\right] v-\tel \frac{|v|}{D}g(\cdot) ~~\ .
\label{eqA:phidot_tauel}
\end{equation}
This model contains the following parameters (in the order they appear above): $\alpha\!\equiv\!\frac{k_B T}{\Omega\sigma_{\hbox{\tiny H}}}$, $b$, $\phi^*$, $\hat{v}\!\equiv\!v_c\,\exp[-\frac{E_0}{k_B T}]$, $D$, $\tilde{f}_0\!\equiv\!\frac{D \mu _0}{h \sigma_{\hbox{\tiny H}}}$ and whatever additional parameters that appear in the threshold function $g(\cdot)$. Note that $\alpha$ defined here is the same one used in Eq.~\eqref{eq:RSF}. The values of these parameters are discussed below in~\hyperref[sec:interfacial elasticity]{Appendix B} and~\hyperref[sec:treadmill]{Appendix C}.

In most of the studies discussed in this work (exceptions are mentioned in Sect.~\ref{sec:linearResponse} and in~\hyperref[sec:interfacial elasticity]{Appendix B}), $g(v)\!=\!\sqrt{1+(v^*/v)^2}$ is used. This choice, which introduces a single additional parameter $v^*$, ensures that $|v|g(v)\!\to\!v^*$ in Eq.~\eqref{eqA:phidot_tauel} is extremely small for $|v|\!\ll\!v^*$, once an extremely small $v^*$ is chosen, and that $|v|g(v)\!\to\!|v|$ for $v\!\gg\!v^*$. For this choice of $g(\cdot)$, the emerging predictions are insensitive to the precise value of $v^*$, as long as it is significantly smaller than all other velocity scales in a given problem. In some applications of the model, most notably when large slip displacements are involved, one assumes that $\tau^{\rm el}$ quickly relaxes to its steady-state. Under these conditions, we set $\dot\tau^{\rm el}\=0$ inside Eq.~\eqref{eqA:phidot_tauel} and substitute the resulting $\tau^{\rm el}$ in Eqs.~\eqref{eqA:RSFsinh}-\eqref{eqA:phidot_tauel} to obtain
\begin{equation}
\label{eqA:RSF_moditfied1}
f(\phi,v)\!=\!\!\left[1\!+\!b\log\left(1+\frac{\phi}{\phi^*}\right)\right]\left[\frac{\tilde{f}_0}{\sgn(v) g(\cdot)} + \alpha\sinh^{-1}\!\!\left(\frac{v}{2\,\hat{v}}\right)\right] ~\qquad\hbox{and}~\qquad
 \dot\phi \!=\! 1- \frac{\phi|v|}{D}g(\cdot)\ .
\end{equation}
The steady-state friction curve $f_{\rm ss}(v)$, which is sketched in Fig.~\ref{fig:ss_friction}, is obtained once we set $\dot\phi\=0$ and substitute the solution for $\phi$ in $f(\phi,v)$ of Eq.~\eqref{eqA:RSF_moditfied1}.

Finally, note that the resulting $f_{\rm ss}(v)$ reduces to the conventional RSF expression in Eq.~\eqref{eq:RSF} once the two constitutive extensions introduced in this paper are removed, i.e.~$\log(1+\phi/\phi^*)\!\to\!\log(\phi/\phi^*)$ and $\mu_0\!\to\!0$ (the latter implies $\tilde{f}_0\=0$) are substituted in Eq.~\eqref{eqA:RSF_moditfied1}, and then the steady-state of $\phi$ is used. To see this, set $g(\cdot)\=1$ ( $g(\cdot)$ accounts for the low-$v$ velocity-strengthening branch in Fig.~\ref{fig:ss_friction}, which is not described by Eq.~\eqref{eq:RSF}), expand $\sinh^{-1}(x)\!\approx\!\log(2x)$ for large argument $x$ and drop the resulting ${\cal O}(\log^2)$ term. The outcome identifies with Eq.~\eqref{eq:RSF}, with the parameters $f_0\!\equiv\!\frac{E_0}{\Omega\,\sigma_{\hbox{\tiny H}}}$ and $\beta\!\equiv\!b f_0$, where $\alpha$ is the same as above (and note that $\hat{v}$ is not used here, i.e.~$\hat{v}\!\equiv\!v_c\,\exp[-\frac{E_0}{k_B T}]$ has been substituted in Eq.~\eqref{eqA:RSF_moditfied1} before $\sinh^{-1}(\cdot)$ has been expanded).

\setcounter{equation}{0}
\renewcommand{\theequation}{B\arabic{equation}}
\subsection{Appendix B: Theoretical predictions for frictional dynamics under small stresses}
\label{sec:interfacial elasticity}

The set of equations used to obtain the results reported on in Fig.~\ref{fig:IE}b is listed and discussed in detail in Sect.~\ref{sec:linearResponse}. The only additional input needed in order to reproduce Fig~\ref{fig:IE}b is the function $g(\cdot)$ and the determination of the material parameters. The experimental data in Fig.~\ref{fig:IE}a clearly indicate that the frictional response up to shear forces of $5\!-\!6$N is predominantly linear elastic, i.e.~that $\tau^{\rm vis}$ is negligible in Eq.~\eqref{eq:KV} (and consequently also in Eq.~\eqref{eqA:RSFsinh}) in this range of loading forces. While the onset of irreversibility is not sharp, it is clear that significant deviations from a linear elastic behavior emerge only for shear forces larger than $7$N. To capture these observations in a phenomenological manner, we set $g(\tau)\=\Theta(\tau-A(\phi)\tau_c)$, where $\Theta$ is the Heaviside step function. Here $\tau_c$ is a characteristic interfacial yield stress, whose value is chosen such that the onset of irreversibility takes place around $7$N (in particular, given $A(\phi(t\=0))$ of Eq.~\eqref{eq:A} as detailed below, we set $\tau_c\=70$MPa). With this choice of $g(\cdot)$, which has also been used in~\cite{Bar-Sinai2012}, the $v\!\approx\!0$ regime in the steady-state friction curve of Fig.~\ref{fig:ss_friction} becomes a strictly vertical (elastic) $v\=0$ line (as shown in~Fig.~1a of~\cite{Bar-Sinai2012}).

The friction parameters for the PMMA plates used in the experiments of~\cite{Berthoud1998}, and which enter Eqs.~\eqref{eqA:RSFsinh}-\eqref{eqA:phidot_tauel}, are the same as those reported on in Table~\ref{AppBTable} below (note that $v^*$ there is irrelevant for the present analysis as $g(\cdot)$ here is characterized by $\tau_c$, as just discussed). The external parameters are as reported explicitly in~\cite{Berthoud1998} (specifically we have $A_n\=0.01$m$^2$, $F_{_{\rm N}}\=24$N and $\mid\!\!dF_{_{\rm S}}/dt\!\!\mid =\!0.5$N/s) and the initial conditions are $\phi(t\=0)\=100$s (which together with $b$ of Table~\ref{AppBTable} allows to determine $A(\phi(t\=0))$ mentioned above), $\dot{\delta}(t\=0)\=0$ and $\tau^{\rm el}(t\=0)\=0$. The only missing parameter is the interfacial elasticity ratio $\mu_0/h$, which is directly extracted from the initial linear slope in the $F_{_{\rm S}}(\delta)$ experimental data (dashed line in Fig.~\ref{fig:IE}a) according to $dF_{_{\rm S}}/\!d\delta\=\frac{\mu_0 F_{_{\rm N}}}{h \sigma_{\hbox{\tiny H}}}[1+b\log(1+\phi(t\=0)/\phi^*)]$, resulting in $\tilde{f}_0\=\frac{D \mu _0}{h \sigma_{\hbox{\tiny H}}}\=0.209$ (which is not the same as the one listed in Table~\ref{AppBTable}, though the two are quite close to each other). Then Eqs.~\eqref{eq:1DOF_F},~\eqref{eqA:RSFsinh} and~\eqref{eqA:phidot_tauel} are solved for $\phi(t)$, $\dot{\delta}(t)$ and $\tau^{\rm el}(t)$ (for the two applied loading protocols $F_{_{\rm S}}(t)$). Finally, $\delta(t)$ of Fig.~\ref{fig:IE}b is obtained (for the two protocols) by integrating $\dot{\delta}(t)$ with $\delta(t\=0)\=0$.

\setcounter{equation}{0}
\renewcommand{\theequation}{C\arabic{equation}}
\subsection{Appendix C: A treadmill FEM routine for obtaining steady-state frictional modes}
\label{sec:treadmill}

The FEM results reported above have been obtained using the FEM software package FreeFem++~\citep{Hecht2012}. The partial differential equations are solved in a monolithic scheme on a triangular mesh using a real space implementation. The fully inertial evolution equations, which include the linear elastodynamic equations (momentum balance with Hooke's law), are expressed in weak form in the spirit of the finite element method. The time derivatives are expressed via finite differences. The frictional boundary condition is implemented via a semi-implicit integration scheme. Details about the results reported on in Fig.~\ref{fig:patch} can be found in~\cite{Aldam2017a}, while here we focus on the results reported on in Fig.~\ref{fig:2D} and Fig.~\ref{fig:spectrum}a.

In these calculations, a deformable body of height $H$ and length $L\!\gg\!H$ sliding on top of a rigid substrate has been considered. A shear stress of magnitude $\tau_0$, which is larger than the minimum of the steady-state friction curve in Fig.~\ref{fig:ss_friction}, and a normal stress $\sigma_{yy}\=-\sigma_0$ are applied at $y\=H$ (cf.~Fig.~\ref{fig:sketch}). The lateral boundaries are traction free and the rigid substrate implies that we have $u_y\=0$ at the interface (which ensures no interpenetration into and detachment from the substrate). In order to efficiently reach steady-state conditions, a ``treadmill'' routine in which the system is shifted such that the propagating mode always remains in the lateral center of the sliding body has been employed. Consequently, we effectively obtain a description in a co-moving reference frame. This requires imposing additional boundary conditions on the fields $\phi$ and $\tau^{\rm el}$ (cf.~Eq.~\eqref{eqA:phidot_tauel}) at the leading edge, which are chosen according to the stable low-velocity fixed-point corresponding to the intersection of $\tau_0$ the steady-state friction curve in Fig.~\ref{fig:ss_friction}. The treadmill scheme also allows to locally increase the spatial resolution near the front position, a main region of interest, without the need for remeshing during propagation, as the front always remains at the same position.

For the largest simulated systems with $H\=16$cm we used a slider of length up to $L\=9.6$m.
Altogether, we typically used non-uniform discretizations with up to $800\!\times\!20$ grid points, and checked that the results are insensitive against further mesh refinement.
To avoid numerical instabilities, the dynamics is suppressed near the trailing and leading edge of the slider, and its length $L$ is chosen to be sufficiently large to avoid their influence on the front dynamics.
We used an initially large timestep, $\Delta t\=10^{-4}$s, which allows for rapid progress in the initial stage of the simulations. This large initial timestep also stabilizes the dynamics, but leads to reduced accuracy. Consequently, we varied the timestep such that it relaxes to a final value of $\Delta t\=10^{-7}$s in the course of the simulations, and we checked that the results are then insensitive against further reduction of the timestep. The simulations typically run until the front has traversed the system length at least once to get rid of sound waves and perturbations which result from the initialization of the system, and which interfere with the front dynamics. After this time, steady-state conditions are reached, and the spatial profiles of all relevant fields are extracted. The results in~Fig.~\ref{fig:2D} and Fig.~\ref{fig:spectrum}a) involve the friction law of Eqs.~\eqref{eqA:RSFsinh}-\eqref{eqA:phidot_tauel} (together with $g(v)\!=\!\sqrt{1+(v^*/v)^2}$), with the bulk and interfacial parameters listed in Table~\ref{AppBTable}. The same interfacial parameters are relevant also for Fig.~\ref{fig:IE}b, see the caption of Table~\ref{AppBTable} and~\hyperref[sec:interfacial elasticity]{Appendix B} for details.

\begin{table}[ht]
  \centering
  \begin{tabular}{|c|c|}
  \hline
  Parameter & Value \\
  \hline
  $\mu$ &  3.1 GPa  \\
$\nu$ & 1/3  \\
$\rho$ & 60 kg/m$^3$ \\
$\alpha$ & $0.005$ \\
$b$ & 0.075 \\
$\phi^*$ & $3.3\cdot10^{-4}$ s \\
$\hat{v}\!=\!v^*$ & $10^{-7}$ m/s \\
$D$ & $0.5\cdot 10^{-6}$ m\\
$\tilde{f}_0$ & $5/18$ \\
$\sigma_{\hbox{\tiny H}}$ & $5.4\cdot10^8$ Pa \\
   \hline
\end{tabular}
  \caption{Bulk and interfacial parameters used in the FEM simulations of Fig.~\ref{fig:2D} and Fig.~\ref{fig:spectrum}a. Note that $\nu$ is Poisson's ratio (not defined in the manuscript), which is needed for Hooke's law, and that an unrealistically small mass density $\rho$ is used (leading to unrealistically large wave speeds), as strongly inertial effects have not been of interest in these particular calculations. The applied normal stress in these calculations has been set to $\sigma_0\!=\!1$MPa. Finally, note that $\sigma_{\hbox{\tiny H}}$ is not needed for solving the equations, just for calculating $A$ of Eq.~\eqref{eq:A}, the result of which is presented in Fig.~\ref{fig:2D}c. The very same interfacial parameter, except for $v^*$ and $\tilde{f}_0$, have been used to generate the results in Fig.~\ref{fig:IE}b (see~\hyperref[sec:interfacial elasticity]{Appendix B} for details).}\label{AppBTable}
\end{table}


\begin{thebibliography}{-------}
\providecommand{\natexlab}[1]{#1}

\bibitem[Bowden and Tabor(1950)]{Bowden1950}
Bowden, F.P.; Tabor, D.
\newblock {\em {The Friction and Lubrication of Solids}}; Clarendon Press,
  1950.

\bibitem[Persson(1998)]{Persson1998}
Persson, B.N.J.
\newblock {\em {Sliding friction: physical principles and applications}};
  Springer Sciense {\&} Buisness Media,  1998.

\bibitem[Svetlizky \em{et~al.}(2019)Svetlizky, Bayart, and
  Fineberg]{Svetlizky2019}
Svetlizky, I.; Bayart, E.; Fineberg, J.
\newblock {Brittle Fracture Theory Describes the Onset of Frictional Motion}.
\newblock {\em Annu. Rev. Condens. Matter Phys.} {\bf 2019}, {\em
  10},~031218--013327.
\newblock
  doi:{\color{black}  \href{https://doi.org/10.1146/annurev-conmatphys-031218-013327}{\detokenize{10.1146/annurev-conmatphys-031218-013327}}}.

\bibitem[S{\o}rensen \em{et~al.}(1996)S{\o}rensen, Jacobsen, and
  Stoltze]{Sorensen1996}
S{\o}rensen, M.; Jacobsen, K.; Stoltze, P.
\newblock {Simulations of atomic-scale sliding friction}.
\newblock {\em Phys. Rev. B} {\bf 1996}, {\em 53},~2101--2113.
\newblock
  doi:{\color{black}\href{https://doi.org/10.1103/PhysRevB.53.2101}{\detokenize{10.1103/PhysRevB.53.2101}}}.

\bibitem[Li \em{et~al.}(2011)Li, Tullis, Goldsby, and Carpick]{Li2011}
Li, Q.; Tullis, T.E.; Goldsby, D.L.; Carpick, R.W.
\newblock {Frictional ageing from interfacial bonding and the origins of rate
  and state friction}.
\newblock {\em Nature} {\bf 2011}, {\em 480},~233--236.
\newblock
  doi:{\color{black}\href{https://doi.org/10.1038/nature10589}{\detokenize{10.1038/nature10589}}}.

\bibitem[Vanossi \em{et~al.}(2013)Vanossi, Manini, Urbakh, Zapperi, and
  Tosatti]{Vanossi2013}
Vanossi, A.; Manini, N.; Urbakh, M.; Zapperi, S.; Tosatti, E.
\newblock {Colloquium: Modeling friction: From nanoscale to mesoscale}.
\newblock {\em Rev. Mod. Phys.} {\bf 2013}, {\em 85},~529--552.
\newblock
  doi:{\color{black}\href{https://doi.org/10.1103/RevModPhys.85.529}{\detokenize{10.1103/RevModPhys.85.529}}}.

\bibitem[Armstrong-H{\'{e}}louvry \em{et~al.}(1994)Armstrong-H{\'{e}}louvry,
  Dupont, and {Canudas de Wit}]{Armstrong-Helouvry1994}
Armstrong-H{\'{e}}louvry, B.; Dupont, P.; {Canudas de Wit}, C.
\newblock {A Survey of Models, Analysis Tools and Compensations Methods for the
  Control of Machines with Friction}.
\newblock {\em Automatica} {\bf 1994}, {\em 30},~1083--1138.
\newblock
  doi:{\color{black}\href{https://doi.org/http://dx.doi.org/10.1016/0005-1098(94)90209-7}{\detokenize{http://dx.doi.org/10.1016/0005-1098(94)90209-7}}}.

\bibitem[Wojewoda \em{et~al.}(2008)Wojewoda, Stefa{\'{n}}ski, Wiercigroch, and
  Kapitaniak]{Wojewoda2008}
Wojewoda, J.; Stefa{\'{n}}ski, A.; Wiercigroch, M.; Kapitaniak, T.
\newblock {Hysteretic effects of dry friction: modelling and experimental
  studies.}
\newblock {\em Philos. Trans. A. Math. Phys. Eng. Sci.} {\bf 2008}, {\em
  366},~747--765.
\newblock
  doi:{\color{black}\href{https://doi.org/10.1098/rsta.2007.2125}{\detokenize{10.1098/rsta.2007.2125}}}.

\bibitem[Marone(1998)]{Marone1998a}
Marone, C.
\newblock {Laboratoty-derived friction laws and their application to seismic
  faulting}.
\newblock {\em Annu. Rev. Earth Planet. Sci.} {\bf 1998}, {\em 26},~643--696.
\newblock
  doi:{\color{black}\href{https://doi.org/10.1146/annurev.earth.26.1.643}{\detokenize{10.1146/annurev.earth.26.1.643}}}.

\bibitem[Ben-Zion(2001)]{Ben-Zion2001}
Ben-Zion, Y.
\newblock {Dynamic ruptures in recent models of earthquake faults}.
\newblock {\em J. Mech. Phys. Solids} {\bf 2001}, {\em 49},~2209--2244.
\newblock
  doi:{\color{black}\href{https://doi.org/10.1016/S0022-5096(01)00036-9}{\detokenize{10.1016/S0022-5096(01)00036-9}}}.

\bibitem[Scholz(2002)]{Scholz2002}
Scholz, C.H.
\newblock {\em {The mechanics of earthquakes and faulting}}; Cambridge
  university press,  2002.

\bibitem[Ohnaka(2013)]{Ohnaka2013}
Ohnaka, M.
\newblock {\em {The physics of rock failure and earthquakes}}; Cambridge
  University Press,  2013.

\bibitem[Gnecco \em{et~al.}(2001)Gnecco, Bennewitz, Gyalog, and
  Meyer]{Gnecco2001}
Gnecco, E.; Bennewitz, R.; Gyalog, T.; Meyer, E.
\newblock {Friction experiments on the nanometre scale}.
\newblock {\em J. Phys. Condens. Matter} {\bf 2001}, {\em 13},~202.
\newblock
  doi:{\color{black}\href{https://doi.org/10.1088/0953-8984/13/31/202}{\detokenize{10.1088/0953-8984/13/31/202}}}.

\bibitem[Baumberger and Caroli(2006)]{Baumberger2006}
Baumberger, T.; Caroli, C.
\newblock {Solid friction from stick–slip down to pinning and aging}.
\newblock {\em Adv. Phys.} {\bf 2006}, {\em 55},~279--348.
\newblock
  doi:{\color{black}\href{https://doi.org/10.1080/00018730600732186}{\detokenize{10.1080/00018730600732186}}}.

\bibitem[Ben-Zion(2008)]{Ben-Zion2008}
Ben-Zion, Y.
\newblock {Collective behavior of earthquakes and faults: Continuum-discrete
  transitions, progressive evolutionary changes, and different dynamic
  regimes}.
\newblock {\em Rev. Geophys.} {\bf 2008}, {\em 46},~RG4006.
\newblock
  doi:{\color{black}\href{https://doi.org/10.1029/2008RG000260}{\detokenize{10.1029/2008RG000260}}}.

\bibitem[Vakis \em{et~al.}(2018)Vakis, Yastrebov, Scheibert, Nicola, Dini,
  Minfray, Almqvist, Paggi, Lee, Limbert, Molinari, Anciaux, Aghababaei,
  {Echeverri Restrepo}, Papangelo, Cammarata, Nicolini, Putignano, Carbone,
  Stupkiewicz, Lengiewicz, Costagliola, Bosia, Guarino, Pugno, M{\"{u}}ser, and
  Ciavarella]{Vakis2018}
Vakis, A.; Yastrebov, V.; Scheibert, J.; Nicola, L.; Dini, D.; Minfray, C.;
  Almqvist, A.; Paggi, M.; Lee, S.; Limbert, G.; Molinari, J.; Anciaux, G.;
  Aghababaei, R.; {Echeverri Restrepo}, S.; Papangelo, A.; Cammarata, A.;
  Nicolini, P.; Putignano, C.; Carbone, G.; Stupkiewicz, S.; Lengiewicz, J.;
  Costagliola, G.; Bosia, F.; Guarino, R.; Pugno, N.; M{\"{u}}ser, M.;
  Ciavarella, M.
\newblock {Modeling and simulation in tribology across scales: An overview}.
\newblock {\em Tribol. Int.} {\bf 2018}, {\em 125},~169--199.
\newblock
  doi:{\color{black}\href{https://doi.org/10.1016/j.triboint.2018.02.005}{\detokenize{10.1016/j.triboint.2018.02.005}}}.

\bibitem[Dieterich(1978)]{Dieterich1978}
Dieterich, J.H.
\newblock {Time-dependent friction and the mechanics of stick-slip}.
\newblock {\em Pure Appl. Geophys.} {\bf 1978}, {\em 116},~790--806.
\newblock
  doi:{\color{black}\href{https://doi.org/10.1007/BF00876539}{\detokenize{10.1007/BF00876539}}}.

\bibitem[Dieterich(1979)]{Dieterich1979}
Dieterich, J.H.
\newblock {Modeling of rock friction: 1. Experimental results and constitutive
  equations}.
\newblock {\em J. Geophys. Res. Solid Earth} {\bf 1979}, {\em 84},~2161.
\newblock
  doi:{\color{black}\href{https://doi.org/10.1029/JB084iB05p02161}{\detokenize{10.1029/JB084iB05p02161}}}.

\bibitem[Rice and Ruina(1983)]{Rice1983}
Rice, J.R.; Ruina, A.L.
\newblock {Stability of Steady Frictional Slipping}.
\newblock {\em J. Appl. Mech.} {\bf 1983}, {\em 50},~343--349.
\newblock
  doi:{\color{black}\href{https://doi.org/10.1115/1.3167042}{\detokenize{10.1115/1.3167042}}}.

\bibitem[Ruina(1983)]{Ruina1983}
Ruina, A.L.
\newblock {Slip instability and state variable friction laws}.
\newblock {\em J. Geophys. Res.} {\bf 1983}, {\em 88},~10359--10370.
\newblock
  doi:{\color{black}\href{https://doi.org/10.1029/JB088iB12p10359}{\detokenize{10.1029/JB088iB12p10359}}}.

\bibitem[Heslot \em{et~al.}(1994)Heslot, Baumberger, Perrin, Caroli, and
  Caroli]{Heslot1994}
Heslot, F.; Baumberger, T.; Perrin, B.; Caroli, B.; Caroli, C.
\newblock {Creep, stick-slip, and dry-friction dynamics: Experiments and a
  heuristic model}.
\newblock {\em Phys. Rev. E} {\bf 1994}, {\em 49},~4973--4988.
\newblock
  doi:{\color{black}\href{https://doi.org/10.1103/PhysRevE.49.4973}{\detokenize{10.1103/PhysRevE.49.4973}}}.

\bibitem[Popov \em{et~al.}(2010)Popov, Grzemba, Starcevic, and
  Fabry]{Popov2010}
Popov, V.L.; Grzemba, B.; Starcevic, J.; Fabry, C.
\newblock {Accelerated creep as a precursor of friction instability and
  earthquake prediction}.
\newblock {\em Phys. Mesomech.} {\bf 2010}, {\em 13},~283--291.
\newblock
  doi:{\color{black}\href{https://doi.org/10.1016/j.physme.2010.11.009}{\detokenize{10.1016/j.physme.2010.11.009}}}.

\bibitem[Dieterich(1986)]{Dieterich1986}
Dieterich, J.H.
\newblock {A model for the nucleation of earthquake slip}. In {\em Earthq.
  source Mech.}; Wiley Online Library,  1986; pp. 37--47.
\newblock
  doi:{\color{black}\href{https://doi.org/10.1029/GM037p0037}{\detokenize{10.1029/GM037p0037}}}.

\bibitem[Rice(1980)]{Rice1980a}
Rice, J.R.
\newblock {The mechanics of earthquake rupture}. In {\em Phys. Earth's Inter.};
   1980; pp. 555--649.
\newblock
  doi:{\color{black}\href{https://doi.org/10.1.1.161.3251}{\detokenize{10.1.1.161.3251}}}.

\bibitem[Rabinowicz(1951)]{Rabinowicz1951}
Rabinowicz, E.
\newblock {The Nature of the Static and Kinetic Coefficients of Friction}.
\newblock {\em J. Appl. Phys.} {\bf 1951}, {\em 22},~1373--1379.
\newblock
  doi:{\color{black}\href{https://doi.org/10.1063/1.1699869}{\detokenize{10.1063/1.1699869}}}.

\bibitem[Dieterich and Kilgore(1994)]{Dieterich1994a}
Dieterich, J.H.; Kilgore, B.D.
\newblock {Direct observation of frictional contacts: New insights for
  state-dependent properties}.
\newblock {\em Pure Appl. Geophys.} {\bf 1994}, {\em 143},~283--302.
\newblock
  doi:{\color{black}\href{https://doi.org/10.1007/BF00874332}{\detokenize{10.1007/BF00874332}}}.

\bibitem[Rubinstein \em{et~al.}(2004)Rubinstein, Cohen, and
  Fineberg]{Rubinstein2004}
Rubinstein, S.M.; Cohen, G.; Fineberg, J.
\newblock {Detachment fronts and the onset of dynamic friction}.
\newblock {\em Nature} {\bf 2004}, {\em 430},~1005--1009.
\newblock
  doi:{\color{black}\href{https://doi.org/10.1038/nature02830}{\detokenize{10.1038/nature02830}}}.

\bibitem[Rubinstein \em{et~al.}(2007)Rubinstein, Cohen, and
  Fineberg]{Rubinstein2007}
Rubinstein, S.M.; Cohen, G.; Fineberg, J.
\newblock {Dynamics of Precursors to Frictional Sliding}.
\newblock {\em Phys. Rev. Lett.} {\bf 2007}, {\em 98},~226103.
\newblock
  doi:{\color{black}\href{https://doi.org/10.1103/PhysRevLett.98.226103}{\detokenize{10.1103/PhysRevLett.98.226103}}}.

\bibitem[Nagata \em{et~al.}(2008)Nagata, Nakatani, and Yoshida]{Nagata2008}
Nagata, K.; Nakatani, M.; Yoshida, S.
\newblock {Monitoring frictional strength with acoustic wave transmission}.
\newblock {\em Geophys. Res. Lett.} {\bf 2008}, {\em 35},~L06310.
\newblock
  doi:{\color{black}\href{https://doi.org/10.1029/2007GL033146}{\detokenize{10.1029/2007GL033146}}}.

\bibitem[Estrin and Br{\'{e}}chet(1996)]{Estrin1996}
Estrin, Y.; Br{\'{e}}chet, Y.
\newblock {On a model of frictional sliding}.
\newblock {\em Pure Appl. Geophys.} {\bf 1996}, {\em 147},~745--762.
\newblock
  doi:{\color{black}\href{https://doi.org/10.1007/BF01089700}{\detokenize{10.1007/BF01089700}}}.

\bibitem[Dieterich(1972)]{Dieterich1972}
Dieterich, J.H.
\newblock {Time-dependent friction in rocks}.
\newblock {\em J. Geophys. Res.} {\bf 1972}, {\em 77},~3690--3697.
\newblock
  doi:{\color{black}\href{https://doi.org/10.1029/JB077i020p03690}{\detokenize{10.1029/JB077i020p03690}}}.

\bibitem[Beeler \em{et~al.}(1994)Beeler, Tullis, and Weeks]{Beeler1994}
Beeler, N.M.; Tullis, T.E.; Weeks, J.D.
\newblock {The roles of time and displacement in the evolution effect in rock
  friction},  1994.
\newblock
  doi:{\color{black}\href{https://doi.org/10.1029/94GL01599}{\detokenize{10.1029/94GL01599}}}.

\bibitem[Berthoud \em{et~al.}(1999)Berthoud, Baumberger, G'Sell, and
  Hiver]{Berthoud1999}
Berthoud, P.; Baumberger, T.; G'Sell, C.; Hiver, J.M.
\newblock {Physical analysis of the state- and rate-dependent friction law:
  Static friction},  1999.
\newblock
  doi:{\color{black}\href{https://doi.org/10.1103/PhysRevB.59.14313}{\detokenize{10.1103/PhysRevB.59.14313}}}.

\bibitem[Bureau \em{et~al.}(2001)Bureau, Baumberger, Caroli, and
  Ronsin]{Bureau2001}
Bureau, L.; Baumberger, T.; Caroli, C.; Ronsin, O.
\newblock {Low-velocity friction between macroscopic solids}.
\newblock {\em Comptes Rendus l'Acad{\'{e}}mie des Sci. - Ser. IV - Phys.} {\bf
  2001}, {\em 2},~699--707.
\newblock
  doi:{\color{black}\href{https://doi.org/10.1016/S1296-2147(01)01212-4}{\detokenize{10.1016/S1296-2147(01)01212-4}}}.

\bibitem[Ben-David \em{et~al.}(2010)Ben-David, Rubinstein, and
  Fineberg]{Ben-David2010}
Ben-David, O.; Rubinstein, S.M.; Fineberg, J.
\newblock {Slip-stick and the evolution of frictional strength.}
\newblock {\em Nature} {\bf 2010}, {\em 463},~76--9.
\newblock
  doi:{\color{black}\href{https://doi.org/10.1038/nature08676}{\detokenize{10.1038/nature08676}}}.

\bibitem[Tullis and Weeks(1986)]{Tullis1986}
Tullis, T.E.; Weeks, J.D.
\newblock {Constitutive behavior and stability of frictional sliding of
  granite}.
\newblock {\em Pure Appl. Geophys.} {\bf 1986}, {\em 124},~383--414.
\newblock
  doi:{\color{black}\href{https://doi.org/10.1007/BF00877209}{\detokenize{10.1007/BF00877209}}}.

\bibitem[Kilgore \em{et~al.}(1993)Kilgore, Blanpied, and
  Dieterich]{Kilgore1993}
Kilgore, B.D.; Blanpied, M.L.; Dieterich, J.H.
\newblock {Velocity dependent friction of granite over a wide range of
  conditions}.
\newblock {\em Geophys. Res. Lett.} {\bf 1993}, {\em 20},~903--906.
\newblock
  doi:{\color{black}\href{https://doi.org/10.1029/93GL00368}{\detokenize{10.1029/93GL00368}}}.

\bibitem[Baumberger and Berthoud(1999)]{Baumberger1999}
Baumberger, T.; Berthoud, P.
\newblock {Physical analysis of the state- and rate-dependent friction law. II.
  Dynamic friction}.
\newblock {\em Phys. Rev. B} {\bf 1999}, {\em 60},~3928--3939.
\newblock
  doi:{\color{black}\href{https://doi.org/10.1103/PhysRevB.60.3928}{\detokenize{10.1103/PhysRevB.60.3928}}}.

\bibitem[Reches and Lockner(2010)]{Reches2010}
Reches, Z.; Lockner, D.A.
\newblock {Fault weakening and earthquake instability by powder lubrication}.
\newblock {\em Nature} {\bf 2010}, {\em 467},~452--455.
\newblock
  doi:{\color{black}\href{https://doi.org/10.1038/nature09348}{\detokenize{10.1038/nature09348}}}.

\bibitem[Rice \em{et~al.}(2001)Rice, Lapusta, and Ranjith]{Rice2001}
Rice, J.R.; Lapusta, N.; Ranjith, K.
\newblock {Rate and state dependent friction and the stability of sliding
  between elastically deformable solids}.
\newblock {\em J. Mech. Phys. Solids} {\bf 2001}, {\em 49},~1865--1898.
\newblock
  doi:{\color{black}\href{https://doi.org/10.1016/S0022-5096(01)00042-4}{\detokenize{10.1016/S0022-5096(01)00042-4}}}.

\bibitem[Nakatani(2001)]{Nakatani2001}
Nakatani, M.
\newblock {Conceptual and physical clarification of rate and state friction:
  Frictional sliding as a thermally activated rheology}.
\newblock {\em J. Geophys. Res. Solid Earth} {\bf 2001}, {\em
  106},~13347--13380.
\newblock
  doi:{\color{black}\href{https://doi.org/10.1029/2000JB900453}{\detokenize{10.1029/2000JB900453}}}.

\bibitem[Beeler \em{et~al.}(2007)Beeler, Tullis, Kronenberg, and
  Reinen]{Beeler2007}
Beeler, N.M.; Tullis, T.E.; Kronenberg, A.K.; Reinen, L.A.
\newblock {The instantaneous rate dependence in low temperature laboratory rock
  friction and rock deformation experiments}.
\newblock {\em J. Geophys. Res.} {\bf 2007}, {\em 112},~B07310.
\newblock
  doi:{\color{black}\href{https://doi.org/10.1029/2005JB003772}{\detokenize{10.1029/2005JB003772}}}.

\bibitem[Bar-Sinai \em{et~al.}(2014)Bar-Sinai, Spatschek, Brener, and
  Bouchbinder]{Bar-Sinai2014}
Bar-Sinai, Y.; Spatschek, R.; Brener, E.A.; Bouchbinder, E.
\newblock {On the velocity-strengthening behavior of dry friction}.
\newblock {\em J. Geophys. Res. Solid Earth} {\bf 2014}, {\em 119},~1738--1748.
\newblock
  doi:{\color{black}\href{https://doi.org/10.1002/2013JB010586}{\detokenize{10.1002/2013JB010586}}}.

\bibitem[Chester and Higgs(1992)]{Chester1992}
Chester, F.M.; Higgs, N.G.
\newblock {Multimechanism friction constitutive model for ultrafine quartz
  gouge at hypocentral conditions}.
\newblock {\em J. Geophys. Res.} {\bf 1992}, {\em 97},~1859--1870.
\newblock
  doi:{\color{black}\href{https://doi.org/10.1029/91JB02349}{\detokenize{10.1029/91JB02349}}}.

\bibitem[Perrin \em{et~al.}(1995)Perrin, Rice, and Zheng]{Perrin1995}
Perrin, G.; Rice, J.R.; Zheng, G.
\newblock {Self-healing slip pulse on a frictional surface}.
\newblock {\em J. Mech. Phys. Solids} {\bf 1995}, {\em 43},~1461--1495.
\newblock
  doi:{\color{black}\href{https://doi.org/10.1016/0022-5096(95)00036-I}{\detokenize{10.1016/0022-5096(95)00036-I}}}.

\bibitem[Putelat and Dawes(2015)]{Putelat2015}
Putelat, T.; Dawes, J.H.
\newblock {Steady and transient sliding under rate-and-state friction}.
\newblock {\em J. Mech. Phys. Solids} {\bf 2015}, {\em 78},~70--93.
\newblock
  doi:{\color{black}\href{https://doi.org/10.1016/j.jmps.2015.01.016}{\detokenize{10.1016/j.jmps.2015.01.016}}}.

\bibitem[Zheng and Rice(1998)]{Zheng1998}
Zheng, G.; Rice, J.R.
\newblock {Conditions under which velocity-weakening friction allows a
  self-healing versus a cracklike mode of rupture}.
\newblock {\em Bull. Seismol. Soc. Am.} {\bf 1998}, {\em 88},~1466--1483.

\bibitem[Courtney-Pratt and Eisner(1957)]{Courtney-Pratt1957}
Courtney-Pratt, J.S.; Eisner, E.
\newblock {The effect of a tangential force on the contact of metallic bodies}.
\newblock {\em Proc. R. Soc. Lond. A.} {\bf 1957}, {\em
  238},~529--550.
\newblock
  doi:{\color{black}\href{https://doi.org/10.1098/rspa.1957.0016}{\detokenize{10.1098/rspa.1957.0016}}}.

\bibitem[Archard(1957)]{Archard1957}
Archard, J.F.
\newblock {Elastic Deformation and the Laws of Friction}.
\newblock {\em Proc. R. Soc. A Math. Phys. Eng. Sci.} {\bf 1957}, {\em
  243},~190--205.
\newblock
  doi:{\color{black}\href{https://doi.org/10.1098/rspa.1957.0214}{\detokenize{10.1098/rspa.1957.0214}}}.

\bibitem[Berthoud and Baumberger(1998)]{Berthoud1998}
Berthoud, P.; Baumberger, T.
\newblock {Shear stiffness of a solid-solid multicontact interface}.
\newblock {\em Proc. R. Soc. A Math. Phys. Eng. Sci.} {\bf 1998}, {\em
  454},~1615--1634.
\newblock
  doi:{\color{black}\href{https://doi.org/10.1098/rspa.1998.0223}{\detokenize{10.1098/rspa.1998.0223}}}.

\bibitem[Bureau \em{et~al.}(2000)Bureau, Baumberger, and Caroli]{Bureau2000}
Bureau, L.; Baumberger, T.; Caroli, C.
\newblock {Shear response of a frictional interface to a normal load
  modulation}.
\newblock {\em Phys. Rev. E} {\bf 2000}, {\em 62},~6810--6820.
\newblock
  doi:{\color{black}\href{https://doi.org/10.1103/PhysRevE.62.6810}{\detokenize{10.1103/PhysRevE.62.6810}}}.

\bibitem[Campa{\~{n}}{\'{a}} \em{et~al.}(2011)Campa{\~{n}}{\'{a}}, Persson, and
  M{\"{u}}ser]{Campana2011}
Campa{\~{n}}{\'{a}}, C.; Persson, B.N.J.; M{\"{u}}ser, M.H.
\newblock {Transverse and normal interfacial stiffness of solids with randomly
  rough surfaces.}
\newblock {\em J. Phys. Condens. Matter} {\bf 2011}, {\em 23},~085001.
\newblock
  doi:{\color{black}\href{https://doi.org/10.1088/0953-8984/23/8/085001}{\detokenize{10.1088/0953-8984/23/8/085001}}}.

\bibitem[Povirk and Needleman(1993)]{Povirk1993}
Povirk, G.L.; Needleman, A.
\newblock {Finite Element Simulations of Fiber Pull-Out}.
\newblock {\em J. Eng. Mater. Technol.} {\bf 1993}, {\em 115},~286.
\newblock
  doi:{\color{black}\href{https://doi.org/10.1115/1.2904220}{\detokenize{10.1115/1.2904220}}}.

\bibitem[Coker \em{et~al.}(2005)Coker, Lykotrafitis, Needleman, and
  Rosakis]{Coker2005a}
Coker, D.; Lykotrafitis, G.; Needleman, A.; Rosakis, A.J.
\newblock {Frictional sliding modes along an interface between identical
  elastic plates subject to shear impact loading}.
\newblock {\em J. Mech. Phys. Solids} {\bf 2005}, {\em 53},~884--922.
\newblock
  doi:{\color{black}\href{https://doi.org/10.1016/j.jmps.2004.11.003}{\detokenize{10.1016/j.jmps.2004.11.003}}}.

\bibitem[Shi \em{et~al.}(2008)Shi, Ben-Zion, and Needleman]{Shi2008}
Shi, Z.; Ben-Zion, Y.; Needleman, A.
\newblock {Properties of dynamic rupture and energy partition in a solid with a
  frictional interface}.
\newblock {\em J. Mech. Phys. Solids} {\bf 2008}, {\em 56},~5--24.
\newblock
  doi:{\color{black}\href{https://doi.org/10.1016/j.jmps.2007.04.006}{\detokenize{10.1016/j.jmps.2007.04.006}}}.

\bibitem[Braun \em{et~al.}(2009)Braun, Barel, and Urbakh]{Braun2009}
Braun, O.M.; Barel, I.; Urbakh, M.
\newblock {Dynamics of Transition from Static to Kinetic Friction}.
\newblock {\em Phys. Rev. Lett.} {\bf 2009}, {\em 103},~194301.
\newblock
  doi:{\color{black}\href{https://doi.org/10.1103/PhysRevLett.103.194301}{\detokenize{10.1103/PhysRevLett.103.194301}}}.

\bibitem[Shi \em{et~al.}(2010)Shi, Needleman, and Ben-Zion]{Shi2010}
Shi, Z.; Needleman, A.; Ben-Zion, Y.
\newblock {Slip modes and partitioning of energy during dynamic frictional
  sliding between identical elastic–viscoplastic solids}.
\newblock {\em Int. J. Fract.} {\bf 2010}, {\em 162},~51--67.
\newblock
  doi:{\color{black}\href{https://doi.org/10.1007/s10704-009-9388-6}{\detokenize{10.1007/s10704-009-9388-6}}}.

\bibitem[Bouchbinder \em{et~al.}(2011)Bouchbinder, Brener, Barel, and
  Urbakh]{Bouchbinder2011}
Bouchbinder, E.; Brener, E.A.; Barel, I.; Urbakh, M.
\newblock {Slow Cracklike Dynamics at the Onset of Frictional Sliding}.
\newblock {\em Phys. Rev. Lett.} {\bf 2011}, {\em 107},~235501.
\newblock
  doi:{\color{black}\href{https://doi.org/10.1103/PhysRevLett.107.235501}{\detokenize{10.1103/PhysRevLett.107.235501}}}.

\bibitem[Rubinstein \em{et~al.}(2011)Rubinstein, Barel, Reches, Braun, Urbakh,
  and Fineberg]{Rubinstein2011}
Rubinstein, S.M.; Barel, I.; Reches, Z.; Braun, O.M.; Urbakh, M.; Fineberg, J.
\newblock {Slip Sequences in Laboratory Experiments Resulting from
  Inhomogeneous Shear as Analogs of Earthquakes Associated with a Fault Edge}.
\newblock {\em Pure Appl. Geophys.} {\bf 2011}, {\em 168},~2151--2166.
\newblock
  doi:{\color{black}\href{https://doi.org/10.1007/s00024-010-0239-1}{\detokenize{10.1007/s00024-010-0239-1}}}.

\bibitem[Bar Sinai \em{et~al.}(2012)Bar Sinai, Brener, and
  Bouchbinder]{Bar-Sinai2012}
Bar Sinai, Y.; Brener, E.A.; Bouchbinder, E.
\newblock {Slow rupture of frictional interfaces}.
\newblock {\em Geophys. Res. Lett.} {\bf 2012}, {\em 39},~L03308.
\newblock
  doi:{\color{black}\href{https://doi.org/10.1029/2011GL050554}{\detokenize{10.1029/2011GL050554}}}.

\bibitem[Bar-Sinai \em{et~al.}(2013)Bar-Sinai, Spatschek, Brener, and
  Bouchbinder]{Bar-Sinai2013}
Bar-Sinai, Y.; Spatschek, R.; Brener, E.A.; Bouchbinder, E.
\newblock {Instabilities at frictional interfaces: Creep patches, nucleation,
  and rupture fronts}.
\newblock {\em Phys. Rev. E} {\bf 2013}, {\em 88},~060403.
\newblock
  doi:{\color{black}\href{https://doi.org/10.1103/PhysRevE.88.060403}{\detokenize{10.1103/PhysRevE.88.060403}}}.

\bibitem[Ferry(1980)]{Ferry1980}
Ferry, J.
\newblock {\em {Viscoelastic Properties of Polymers, 3rd Edition}}; Wiley,
  1980.

\bibitem[Thomson(1865)]{Thomson1865}
Thomson, W.
\newblock {On the Elasticity and Viscosity of Metals}.
\newblock {\em Proc. R. Soc. London} {\bf 1865}, {\em 14},~289--297.
\newblock
  doi:{\color{black}\href{https://doi.org/10.1098/rspl.1865.0052}{\detokenize{10.1098/rspl.1865.0052}}}.

\bibitem[Bureau \em{et~al.}(2003)Bureau, Caroli, and Baumberger]{Bureau2003}
Bureau, L.; Caroli, C.; Baumberger, T.
\newblock {Elasticity and onset of frictional dissipation at a non-sliding
  multi-contact interface}.
\newblock {\em Proc. R. Soc. A Math. Phys. Eng. Sci.} {\bf 2003}, {\em
  459},~2787--2805.
\newblock
  doi:{\color{black}\href{https://doi.org/10.1098/rspa.2003.1146}{\detokenize{10.1098/rspa.2003.1146}}}.

\bibitem[Nakatani and Scholz(2006)]{Nakatani2006}
Nakatani, M.; Scholz, C.H.
\newblock {Intrinsic and apparent short-time limits for fault healing: Theory,
  observations, and implications for velocity-dependent friction}.
\newblock {\em J. Geophys. Res. Solid Earth} {\bf 2006}, {\em 111},~1--19.
\newblock
  doi:{\color{black}\href{https://doi.org/10.1029/2005JB004096}{\detokenize{10.1029/2005JB004096}}}.

\bibitem[Nagata \em{et~al.}(2012)Nagata, Nakatani, and Yoshida]{Nagata2012}
Nagata, K.; Nakatani, M.; Yoshida, S.
\newblock {A revised rate- and state-dependent friction law obtained by
  constraining constitutive and evolution laws separately with laboratory
  data}.
\newblock {\em J. Geophys. Res. Solid Earth} {\bf 2012}, {\em 117},~B02314.
\newblock
  doi:{\color{black}\href{https://doi.org/10.1029/2011JB008818}{\detokenize{10.1029/2011JB008818}}}.

\bibitem[Marone(1998)]{Marone1998}
Marone, C.
\newblock {The effect of loading rate on static friction and the rate of fault
  healing during the earthquake cycle}.
\newblock {\em Nature} {\bf 1998}, {\em 391},~69--72.
\newblock
  doi:{\color{black}\href{https://doi.org/10.1038/34157}{\detokenize{10.1038/34157}}}.

\bibitem[Shimamoto(1986)]{Shimamoto1986}
Shimamoto, T.
\newblock {Transition Between Frictional Slip and Ductile Flow for Halite Shear
  Zones at Room Temperature}.
\newblock {\em Science} {\bf 1986}, {\em 231},~711--714.
\newblock
  doi:{\color{black}\href{https://doi.org/10.1126/science.231.4739.711}{\detokenize{10.1126/science.231.4739.711}}}.

\bibitem[Bar-Sinai \em{et~al.}(2015)Bar-Sinai, Spatschek, Brener, and
  Bouchbinder]{Bar-Sinai2015a}
Bar-Sinai, Y.; Spatschek, R.; Brener, E.A.; Bouchbinder, E.
\newblock {Velocity-strengthening friction significantly affects interfacial
  dynamics, strength and dissipation}.
\newblock {\em Sci. Rep.} {\bf 2015}, {\em 5},~7841.
\newblock
  doi:{\color{black}\href{https://doi.org/10.1038/srep07841}{\detokenize{10.1038/srep07841}}}.

\bibitem[{Di Toro} \em{et~al.}(2004){Di Toro}, Goldsby, and Tullis]{DiToro2004}
{Di Toro}, G.; Goldsby, D.L.; Tullis, T.E.
\newblock {Friction falls towards zero in quartz rock as slip velocity
  approaches seismic rates}.
\newblock {\em Nature} {\bf 2004}, {\em 427},~436--439.
\newblock
  doi:{\color{black}\href{https://doi.org/10.1038/nature02249}{\detokenize{10.1038/nature02249}}}.

\bibitem[Goldsby and Tullis(2011)]{Goldsby2011}
Goldsby, D.L.; Tullis, T.E.
\newblock {Flash Heating Leads to Low Frictional Strength of Crustal Rocks at
  Earthquake Slip Rates}.
\newblock {\em Science} {\bf 2011}, {\em 334},~216--218.
\newblock
  doi:{\color{black}\href{https://doi.org/10.1126/science.1207902}{\detokenize{10.1126/science.1207902}}}.

\bibitem[Weertman(1980)]{Weertman1980}
Weertman, J.
\newblock {Unstable slippage across a fault that separates elastic media of
  different elastic constants}.
\newblock {\em J. Geophys. Res. Solid Earth} {\bf 1980}, {\em 85},~1455--1461.
\newblock
  doi:{\color{black}\href{https://doi.org/10.1029/JB085iB03p01455}{\detokenize{10.1029/JB085iB03p01455}}}.

\bibitem[Aldam \em{et~al.}(2016)Aldam, Bar-Sinai, Svetlizky, Brener, Fineberg,
  and Bouchbinder]{Aldam2016}
Aldam, M.; Bar-Sinai, Y.; Svetlizky, I.; Brener, E.A.; Fineberg, J.;
  Bouchbinder, E.
\newblock {Frictional Sliding without Geometrical Reflection Symmetry}.
\newblock {\em Phys. Rev. X} {\bf 2016}, {\em 6},~041023.
\newblock
  doi:{\color{black}\href{https://doi.org/10.1103/PhysRevX.6.041023}{\detokenize{10.1103/PhysRevX.6.041023}}}.

\bibitem[Heimisson \em{et~al.}(2019)Heimisson, Dunham, and
  Almquist]{Heimisson2019}
Heimisson, E.R.; Dunham, E.M.; Almquist, M.
\newblock {Poroelastic effects destabilize mildly rate-strengthening friction
  to generate stable slow slip pulses}.
\newblock {\em J. Mech. Phys. Solids} {\bf 2019}, {\em 130},~262--279.
\newblock
  doi:{\color{black}\href{https://doi.org/10.1016/j.jmps.2019.06.007}{\detokenize{10.1016/j.jmps.2019.06.007}}}.

\bibitem[Geubelle and Rice(1995)]{Geubelle1995}
Geubelle, P.; Rice, J.R.
\newblock {A spectral method for three-dimensional elastodynamic fracture
  problems}.
\newblock {\em J. Mech. Phys. Solids} {\bf 1995}, {\em 43},~1791--1824.
\newblock
  doi:{\color{black}\href{https://doi.org/10.1016/0022-5096(95)00043-I}{\detokenize{10.1016/0022-5096(95)00043-I}}}.

\bibitem[Morrissey and Geubelle(1997)]{Morrissey1997}
Morrissey, J.W.; Geubelle, P.H.
\newblock {A numerical scheme for mode III dynamic fracture problems}.
\newblock {\em Int. J. Numer. Methods Eng.} {\bf 1997}, {\em 40},~1181--1196.
\newblock
  doi:{\color{black}\href{https://doi.org/10.1002/(SICI)1097-0207(19970415)40:7<1181::AID-NME108>3.0.CO;2-X}{\detokenize{10.1002/(SICI)1097-0207(19970415)40:7<1181::AID-NME108>3.0.CO;2-X}}}.

\bibitem[Breitenfeld and Geubelle(1998)]{Breitenfeld1998}
Breitenfeld, M.S.; Geubelle, P.H.
\newblock {Numerical analysis of dynamic debonding under 2D in-plane and 3D
  loading}.
\newblock {\em Int. J. Fract.} {\bf 1998}, {\em 93},~13--38.
\newblock
  doi:{\color{black}\href{https://doi.org/10.1023/A:1007535703095}{\detokenize{10.1023/A:1007535703095}}}.

\bibitem[Ben-Zion and Rice(1995)]{Ben-Zion1995}
Ben-Zion, Y.; Rice, J.R.
\newblock {Slip patterns and earthquake populations along different classes of
  faults in elastic solids}.
\newblock {\em J. Geophys. Res. Solid Earth} {\bf 1995}, {\em
  100},~12959--12983.
\newblock
  doi:{\color{black}\href{https://doi.org/10.1029/94JB03037}{\detokenize{10.1029/94JB03037}}}.

\bibitem[Crupi and Bizzarri(2013)]{Crupi2013}
Crupi, P.; Bizzarri, A.
\newblock {The role of radiation damping in the modeling of repeated earthquake
  events}.
\newblock {\em Ann. Geophys.} {\bf 2013}, {\em 56},~R0111.
\newblock
  doi:{\color{black}\href{https://doi.org/10.4401/ag-6200}{\detokenize{10.4401/ag-6200}}}.

\bibitem[Weertman(1965)]{Weertman1965}
Weertman, J.
\newblock Relationship between displacements on a free surface and the stress
  on a fault.
\newblock {\em Bulletin of the Seismological Society of America} {\bf 1965},
  {\em 55},~945--953.

\bibitem[Acheson(1990)]{Acheson1990}
Acheson, D.J.
\newblock {\em {Elementary fluid dynamics}}; Oxford University Press,  1990.

\bibitem[Viesca(2016)]{Viesca2016b}
Viesca, R.C.
\newblock {Self-similar slip instability on interfaces with rate- and
  state-dependent friction}.
\newblock {\em Proc. R. Soc. A Math. Phys. Eng. Sci.} {\bf 2016}, {\em
  472},~20160254.
\newblock
  doi:{\color{black}\href{https://doi.org/10.1098/rspa.2016.0254}{\detokenize{10.1098/rspa.2016.0254}}}.

\bibitem[Thomsen(1999)]{Thomsen1999}
Thomsen, J.
\newblock {Using fast vibrations to quench friction-induced oscillations}.
\newblock {\em J. Sound Vib.} {\bf 1999}, {\em 228},~1079--1102.
\newblock
  doi:{\color{black}\href{https://doi.org/10.1006/jsvi.1999.2460}{\detokenize{10.1006/jsvi.1999.2460}}}.

\bibitem[Rhee \em{et~al.}(1991)Rhee, Jacko, and Tsang]{Rhee1991}
Rhee, S.; Jacko, M.; Tsang, P.
\newblock {The role of friction film in friction, wear and noise of automotive
  brakes}.
\newblock {\em Wear} {\bf 1991}, {\em 146},~89--97.
\newblock
  doi:{\color{black}\href{https://doi.org/10.1016/0043-1648(91)90226-K}{\detokenize{10.1016/0043-1648(91)90226-K}}}.

\bibitem[Brace and Byerlee(1966)]{Brace1966}
Brace, W.F.; Byerlee, J.D.
\newblock {Stick-Slip as a Mechanism for Earthquakes}.
\newblock {\em Science} {\bf 1966}, {\em 153},~990--992.
\newblock
  doi:{\color{black}\href{https://doi.org/10.1126/science.153.3739.990}{\detokenize{10.1126/science.153.3739.990}}}.

\bibitem[Dieterich(1975)]{Dieterich1975}
Dieterich, J.H.
\newblock {Model for earthquake precursers based on premonitory fault slip}.
\newblock {\em Trans. Geophys. Union} {\bf 1975}, {\em 56},~1059--1060.

\bibitem[Aldam \em{et~al.}(2017)Aldam, Weikamp, Spatschek, Brener, and
  Bouchbinder]{Aldam2017a}
Aldam, M.; Weikamp, M.; Spatschek, R.; Brener, E.A.; Bouchbinder, E.
\newblock {Critical Nucleation Length for Accelerating Frictional Slip}.
\newblock {\em Geophys. Res. Lett.} {\bf 2017}, {\em 44},~11,390--11,398.
\newblock
  doi:{\color{black}\href{https://doi.org/10.1002/2017GL074939}{\detokenize{10.1002/2017GL074939}}}.

\bibitem[Dieterich(1992)]{Dieterich1992}
Dieterich, J.H.
\newblock {Earthquake nucleation on faults with rate-and state-dependent
  strength}.
\newblock {\em Tectonophysics} {\bf 1992}, {\em 211},~115--134.
\newblock
  doi:{\color{black}\href{https://doi.org/10.1016/0040-1951(92)90055-B}{\detokenize{10.1016/0040-1951(92)90055-B}}}.

\bibitem[Brener \em{et~al.}(2016)Brener, Weikamp, Spatschek, Bar-Sinai, and
  Bouchbinder]{Brener2016}
Brener, E.A.; Weikamp, M.; Spatschek, R.; Bar-Sinai, Y.; Bouchbinder, E.
\newblock {Dynamic instabilities of frictional sliding at a bimaterial
  interface}.
\newblock {\em J. Mech. Phys. Solids} {\bf 2016}, {\em 89},~149--173.
\newblock
  doi:{\color{black}\href{https://doi.org/10.1016/j.jmps.2016.01.009}{\detokenize{10.1016/j.jmps.2016.01.009}}}.

\bibitem[Lapusta and Rice(2003)]{Lapusta2003}
Lapusta, N.; Rice, J.R.
\newblock {Nucleation and early seismic propagation of small and large events
  in a crustal earthquake model}.
\newblock {\em J. Geophys. Res. Solid Earth} {\bf 2003}, {\em 108},~2205.
\newblock
  doi:{\color{black}\href{https://doi.org/10.1029/2001JB000793}{\detokenize{10.1029/2001JB000793}}}.

\bibitem[Rubin and Ampuero(2005)]{Rubin2005}
Rubin, A.M.; Ampuero, J.P.
\newblock {Earthquake nucleation on (aging) rate and state faults}.
\newblock {\em J. Geophys. Res. Solid Earth} {\bf 2005}, {\em 110},~B11312.
\newblock
  doi:{\color{black}\href{https://doi.org/10.1029/2005JB003686}{\detokenize{10.1029/2005JB003686}}}.

\bibitem[Ampuero and Rubin(2008)]{Ampuero2008}
Ampuero, J.P.; Rubin, A.M.
\newblock {Earthquake nucleation on rate and state faults – Aging and slip
  laws}.
\newblock {\em J. Geophys. Res. Solid Earth} {\bf 2008}, {\em 113},~B01302.
\newblock
  doi:{\color{black}\href{https://doi.org/10.1029/2007JB005082}{\detokenize{10.1029/2007JB005082}}}.

\bibitem[Kammer \em{et~al.}(2012)Kammer, Yastrebov, Spijker, and
  Molinari]{Kammer2012}
Kammer, D.S.; Yastrebov, V.A.; Spijker, P.; Molinari, J.F.
\newblock {On the Propagation of Slip Fronts at Frictional Interfaces}.
\newblock {\em Tribol. Lett.} {\bf 2012}, {\em 48},~27--32.
\newblock
  doi:{\color{black}\href{https://doi.org/10.1007/s11249-012-9920-0}{\detokenize{10.1007/s11249-012-9920-0}}}.

\bibitem[Svetlizky and Fineberg(2014)]{Svetlizky2014}
Svetlizky, I.; Fineberg, J.
\newblock {Classical shear cracks drive the onset of dry frictional motion}.
\newblock {\em Nature} {\bf 2014}, {\em 509},~205--208.
\newblock
  doi:{\color{black}\href{https://doi.org/10.1038/nature13202}{\detokenize{10.1038/nature13202}}}.

\bibitem[Brener \em{et~al.}(2005)Brener, Malinin, and Marchenko]{Brener2005}
Brener, E.A.; Malinin, S.V.; Marchenko, V.I.
\newblock {Fracture and friction: Stick-slip motion}.
\newblock {\em Eur. Phys. J. E} {\bf 2005}, {\em 17},~101--113.
\newblock
  doi:{\color{black}\href{https://doi.org/10.1140/epje/i2004-10112-3}{\detokenize{10.1140/epje/i2004-10112-3}}}.

\bibitem[Cocco and Bizzarri(2002)]{Cocco2002}
Cocco, M.; Bizzarri, A.
\newblock {On the slip-weakening behavior of rate- and state dependent
  constitutive laws}.
\newblock {\em Geophys. Res. Lett.} {\bf 2002}, {\em 29},~1516.
\newblock
  doi:{\color{black}\href{https://doi.org/10.1029/2001GL013999}{\detokenize{10.1029/2001GL013999}}}.

\bibitem[Freund(1998)]{Freund1998}
Freund, L.B.
\newblock {\em {Dynamic Fracture Mechanics}}; Cambridge university press:
  Cambridge,  1998.

\bibitem[Ben-David \em{et~al.}(2010)Ben-David, Cohen, and
  Fineberg]{Ben-David2010a}
Ben-David, O.; Cohen, G.; Fineberg, J.
\newblock {The Dynamics of the Onset of Frictional Slip}.
\newblock {\em Science} {\bf 2010}, {\em 330},~211--214.
\newblock
  doi:{\color{black}\href{https://doi.org/10.1126/science.1194777}{\detokenize{10.1126/science.1194777}}}.

\bibitem[Kaproth and Marone(2013)]{Kaproth2013}
Kaproth, B.M.; Marone, C.
\newblock {Slow Earthquakes, Preseismic Velocity Changes, and the Origin of
  Slow Frictional Stick-Slip}.
\newblock {\em Science} {\bf 2013}, {\em 341},~1229--1232.
\newblock
  doi:{\color{black}\href{https://doi.org/10.1126/science.1239577}{\detokenize{10.1126/science.1239577}}}.

\bibitem[B{\"{u}}rgmann(2018)]{Burgmann2018}
B{\"{u}}rgmann, R.
\newblock {The geophysics, geology and mechanics of slow fault slip}.
\newblock {\em Earth Planet. Sci. Lett.} {\bf 2018}, {\em 495},~112--134.
\newblock
  doi:{\color{black}\href{https://doi.org/10.1016/j.epsl.2018.04.062}{\detokenize{10.1016/j.epsl.2018.04.062}}}.


\bibitem[Putelat \em{et~al.}(2017)Putelat, Dawes, and Champneys]{Putelat2017}
Putelat, T.; Dawes, J.H.; Champneys, A.R.
\newblock {A phase-plane analysis of localized frictional waves}.
\newblock {\em Proc. R. Soc. A Math. Phys. Eng. Sci.} {\bf 2017}, {\em
  473},~20160606.
\newblock
  doi:{\color{black}\href{https://doi.org/10.1098/rspa.2016.0606}{\detokenize{10.1098/rspa.2016.0606}}}.

\bibitem[Brener \em{et~al.}(2018)Brener, Aldam, Barras, Molinari, and
  Bouchbinder]{Brener2018}
Brener, E.A.; Aldam, M.; Barras, F.; Molinari, J.F.; Bouchbinder, E.
\newblock {Unstable Slip Pulses and Earthquake Nucleation as a Nonequilibrium
  First-Order Phase Transition}.
\newblock {\em Phys. Rev. Lett.} {\bf 2018}, {\em 121},~234302.
\newblock
  doi:{\color{black}\href{https://doi.org/10.1103/PhysRevLett.121.234302}{\detokenize{10.1103/PhysRevLett.121.234302}}}.

\bibitem[Heaton(1990)]{Heaton1990}
Heaton, T.H.
\newblock {Evidence for and implications of self-healing pulses of slip in
  earthquake rupture}.
\newblock {\em Phys. Earth Planet. Inter.} {\bf 1990}, {\em 64},~1--20.
\newblock
  doi:{\color{black}\href{https://doi.org/10.1016/0031-9201(90)90002-F}{\detokenize{10.1016/0031-9201(90)90002-F}}}.

\bibitem[Noda \em{et~al.}(2009)Noda, Dunham, and Rice]{Noda2009}
Noda, H.; Dunham, E.M.; Rice, J.R.
\newblock {Earthquake ruptures with thermal weakening and the operation of
  major faults at low overall stress levels}.
\newblock {\em J. Geophys. Res. Solid Earth} {\bf 2009}, {\em 114},~B07302.
\newblock
  doi:{\color{black}\href{https://doi.org/10.1029/2008JB006143}{\detokenize{10.1029/2008JB006143}}}.

\bibitem[Uenishi and Rice(2003)]{Uenishi2003}
Uenishi, K.; Rice, J.R.
\newblock {Universal nucleation length for slip-weakening rupture instability
  under nonuniform fault loading}.
\newblock {\em J. Geophys. Res. Solid Earth} {\bf 2003}, {\em 108},~2042.
\newblock
  doi:{\color{black}\href{https://doi.org/10.1029/2001JB001681}{\detokenize{10.1029/2001JB001681}}}.

\bibitem[Barras \em{et~al.}(2019{\natexlab{a}})Barras, Aldam, Roch, Brener,
  Bouchbinder, and Molinari]{PartI}
Barras, F.; Aldam, M.; Roch, T.; Brener, E.A.; Bouchbinder, E.; Molinari, J.F.
\newblock {The emergence of crack-like behavior of frictional rupture: The
  origin of stress drops}.
\newblock {\em arXiv}  {\bf 2019}, {\em arXiv:1906.11533}.
\newblock  \href{https://arxiv.org/abs/1906.11533}{{\normalfont
  [1906.11533]}}.

\bibitem[Barras \em{et~al.}(2019{\natexlab{b}})Barras, Aldam, Roch, Brener,
  Bouchbinder, and Molinari]{PartII}
Barras, F.; Aldam, M.; Roch, T.; Brener, E.A.; Bouchbinder, E.; Molinari, J.F.
\newblock {The emergence of crack-like behavior of frictional rupture: Edge
  singularity and energy balance}.
\newblock {\em arXiv}  {\bf 2019}, {\em arXiv:1907.04376}.
\newblock  \href{https://arxiv.org/abs/1907.04376}{{\normalfont
  [1907.04376]}}.

\bibitem[Lu \em{et~al.}(2010)Lu, Rosakis, and Lapusta]{Lu2010}
Lu, X.; Rosakis, A.J.; Lapusta, N.
\newblock {Rupture modes in laboratory earthquakes: Effect of fault prestress
  and nucleation conditions}.
\newblock {\em J. Geophys. Res. Solid Earth} {\bf 2010}, {\em 115},~1--25.
\newblock
  doi:{\color{black}\href{https://doi.org/10.1029/2009JB006833}{\detokenize{10.1029/2009JB006833}}}.

\bibitem[Ida(1972)]{Ida1972}
Ida, Y.
\newblock {Cohesive force across the tip of a longitudinal-shear crack and
  Griffith's specific surface energy}.
\newblock {\em J. Geophys. Res.} {\bf 1972}, {\em 77},~3796--3805.
\newblock
  doi:{\color{black}\href{https://doi.org/10.1029/JB077i020p03796}{\detokenize{10.1029/JB077i020p03796}}}.

\bibitem[Palmer and Rice(1973)]{Palmer1973}
Palmer, A.C.; Rice, J.R.
\newblock {The growth of slip surfaces in the progressive failure of
  over-consolidated clay}.
\newblock {\em Proc. R. Soc. A Math. Phys. Eng. Sci.} {\bf 1973}, {\em
  332},~527--548.
\newblock
  doi:{\color{black}\href{https://doi.org/10.1098/rspa.1973.0040}{\detokenize{10.1098/rspa.1973.0040}}}.

\bibitem[Woodhouse \em{et~al.}(2015)Woodhouse, Putelat, and
  McKay]{Woodhouse2015}
Woodhouse, J.; Putelat, T.; McKay, A.
\newblock {Are there reliable constitutive laws for dynamic friction?}
\newblock {\em Philos. Trans. R. Soc. A Math. Phys. Eng. Sci.} {\bf 2015}, {\em
  373},~20140401.
\newblock
  doi:{\color{black}\href{https://doi.org/10.1098/rsta.2014.0401}{\detokenize{10.1098/rsta.2014.0401}}}.

\bibitem[Radiguet \em{et~al.}(2013)Radiguet, Kammer, Gillet, and
  Molinari]{Radiguet2013}
Radiguet, M.; Kammer, D.S.; Gillet, P.; Molinari, J.F.
\newblock {Survival of Heterogeneous Stress Distributions Created by Precursory
  Slip at Frictional Interfaces}.
\newblock {\em Phys. Rev. Lett.} {\bf 2013}, {\em 111},~164302.
\newblock
  doi:{\color{black}\href{https://doi.org/10.1103/PhysRevLett.111.164302}{\detokenize{10.1103/PhysRevLett.111.164302}}}.

\bibitem[Radiguet \em{et~al.}(2015)Radiguet, Kammer, and
  Molinari]{Radiguet2015}
Radiguet, M.; Kammer, D.S.; Molinari, J.F.
\newblock {The role of viscoelasticity on heterogeneous stress fields at
  frictional interfaces}.
\newblock {\em Mech. Mater.} {\bf 2015}, {\em 80},~276--287.
\newblock
  doi:{\color{black}\href{https://doi.org/10.1016/j.mechmat.2014.03.009}{\detokenize{10.1016/j.mechmat.2014.03.009}}}.

\bibitem[Allison and Dunham(2017)]{Allison2017}
Allison, K.L.; Dunham, E.M.
\newblock {Earthquake cycle simulations with rate-and-state friction and
  power-law viscoelasticity}.
\newblock {\em Tectonophysics} {\bf 2017}.
\newblock
  doi:{\color{black}\href{https://doi.org/10.1016/j.tecto.2017.10.021}{\detokenize{10.1016/j.tecto.2017.10.021}}}.

\bibitem[Dunham and Rice(2008)]{Dunham2008}
Dunham, E.M.; Rice, J.R.
\newblock {Earthquake slip between dissimilar poroelastic materials}.
\newblock {\em J. Geophys. Res. Solid Earth} {\bf 2008}, {\em 113},~B09304.
\newblock
  doi:{\color{black}\href{https://doi.org/10.1029/2007JB005405}{\detokenize{10.1029/2007JB005405}}}.

\bibitem[Hecht(2012)]{Hecht2012}
Hecht, F.
\newblock {New development in freefem++}.
\newblock {\em J. Numer. Math.} {\bf 2012}, {\em 20}.
\newblock
  doi:{\color{black}\href{https://doi.org/10.1515/jnum-2012-0013}{\detokenize{10.1515/jnum-2012-0013}}}.

\end{thebibliography}

\end{document}